\documentclass[prd,showpacs,preprintnumbers,amsmath,amssymb,floatfix]{revtex4}
\usepackage{graphics}
\usepackage{epsfig}
\usepackage{subfigure}
\usepackage{dcolumn}% Align table columns on decimal point
\usepackage{bm}% bold math
\usepackage{color}

\begin{document}
\draft
\title{{\bf Bound States of the Heavy Flavor Vector Mesons and $Y(4008)$ and $Z^{+}_1(4050)$  }}

\author{Gui-Jun Ding}

\affiliation{\centerline{Department of Modern
Physics,}\centerline{University of Science and Technology of
China,Hefei, Anhui 230026, China}}

\begin{abstract}
The $D^{*}\bar{D}^{*}$ and $B^{*}\bar{B}^{*}$ systems are studied
dynamically in the one boson exchange model, where $\pi$, $\eta$,
$\sigma$, $\rho$ and $\omega$ exchanges are taken into account. Ten
allowed states with low spin parity are considered. We suggest that
the $1^{--}$, $2^{++}$, $0^{++}$ and $0^{-+}$ $B^{*}\bar{B}^{*}$
molecules should exist, and the $D^{*}\bar{D}^{*}$ bound states with
the same quantum numbers very likely exist as well. However, the CP
exotic ($1^{-+}$, $2^{+-}$) $B^{*}\bar{B}^{*}$ and
$D^{*}\bar{D}^{*}$ states may not be bound by the one boson exchange
potential. We find that the $I=0$ configuration is more deeply bound
than the $I=1$ configuration, hence $Z^{+}_1(4050)$ may not be a
$D^{*}\bar{D}^{*}$ molecule. Although $Y(4008)$ is close to the
$D^{*}\bar{D}^{*}$ threshold, the interpretation of $Y(4008)$ as a
$D^{*}\bar{D}^{*}$ molecule is not favored by its huge width.
$1^{--}$ $D^{*}\bar{D}^{*}$ and $B^{*}\bar{B}^{*}$ states can be
produced copiously in $e^{+}e^{-}$ annihilation, detailed scanning
of the $e^{+}e^{-}$ annihilation data near the $D^{*}\bar{D}^{*}$
and $B^{*}\bar{B}^{*}$ threshold is an important check to our
predictions.

\end{abstract}

\maketitle

\vskip0.5cm

\section{Introduction}

In the past years, the observations of a number of charmonium-like
"$X, Y, Z$" mesons at $B$ factories have stimulated the interest in
the spectroscopy of the charmonium states again. There is growing
evidence that at least some of these new states are non-conventional
$c\bar{c}$ states, such as deuteron like hadronic molecules,
tetraquark states or hybrid have been suggested \cite{review}. Among
these new mesons, some are very close to the threshold of two
charmed mesons, such as $X(3872)$ \cite{Choi:2003ue} and
$Z^{+}(4430)$ \cite{Choi:2007wga,Aubert:2008nk}. This distinctive
character inspires the molecular interpretation for these mesons. In
particular, some new enhancements near the $D^{*}\bar{D}^{*}$
threshold have been observed recently.

The Belle collaboration reported a broad $\pi^{+}\pi^{-}J/\psi$ peak
near 4008 MeV in addition to the well-known state $Y(4260)$ by
studying the initial state radiation process
$e^{+}e^{-}\rightarrow\gamma_{ISR}\pi^{+}\pi^{-}J/\psi$
\cite{Yuan:2007sj}, its mass and width are fitted to be
$M=(4008\pm40^{+114}_{-28})$ MeV and $\Gamma=(226\pm44\pm87)$ MeV
respectively. We notice its width is huge. This peak was suggested
to be related to the $D^{*}\bar{D}^{*}$ threshold and could be a
$D^{*}\bar{D}^{*}$ molecule in Ref. \cite{Liu:2007ez}. The
$\pi^{+}\pi^{-}J/\psi$ spectrum was studied further by the Babar
collaboration. However, there was no evidence for this broad
enhancement, and the an upper limit ${\cal
B}(\pi^{+}\pi^{-}J/\psi)\Gamma_{e^{+}e^{-}}<0.7 eV$ at $90\%$ C.L.
was obtained \cite{Aubert:2008ic}. $Y(4008)$ is far from being
established so far, more experimental efforts are obviously needed.

Of special importance is the observation of the state carrying
non-zero electric charge with hidden charm quarks. Since the
observation of $Z^{+}(4430)$ \cite{Choi:2007wga,Aubert:2008nk}, the
Belle collaboration reported two resonance-like structures
$Z^{+}_1(4050)$ and $Z^{+}_2(4250)$ in the $\pi^{+}\chi_{c1}$ mass
distribution in the exclusive process $\bar{B}^{0}\rightarrow
K^{-}\pi^{+}\chi_{c1}$ \cite{Mizuk:2008me}. Their masses and widths
are determined to be $M_1=(4051\pm14^{+20}_{-41})$ MeV,
$\Gamma_1=82^{+21+47}_{-17-22}$ MeV, $M_2=(4248^{+44+180}_{-29-35})$
MeV and $\Gamma_2=(177^{+54+316}_{-39-61})$ MeV respectively. Since
$\pi^{+}$ is an isovector with negative G-parity, and $\chi_{c1}$ is
a isospin singlet with positive G-parity, the quantum numbers of
both Z$_1^{+}$(4051) and Z$_2^{+}$(4250) are ${I^{G}=1^{-}}$. In
Ref.\cite{Liu:2008tn}, $Z^{+}_1(4050)$ was suggested to be possibly
a $J^{P}=0^{+}$ $D^{*}\bar{D}^{*}$ molecule due to its closeness to
the $D^{*}\bar{D}^{*}$ threshold. However, the QCD sum rule results
indicated that the $D^{*+}\bar{D}^{*0}$ state is probably a virtual
state, which is not related with the $Z^{+}_1(4050)$ resonance-like
structure \cite{Lee:2008gn}. In addition, we demonstrated that
$Z^{+}_2(4250)$ as a ${D_1\bar{D}/D\bar{D}_1}$ or
${D_0\bar{D}^{*}/D^{*}\bar{D}_0}$ molecule is disfavored in Ref.
\cite{Ding:2008gr}.

Since the repulsive kinetic energy is greatly reduced by the heavy
quark mass, the interaction between light quarks is strong enough so
that the molecular states consisting of heavy flavor mesons very
likely exist. In fact, the hadronic molecules consisting of two
charm mesons were suggested long ago \cite{Voloshin:1976ap}. De
Rujula, Georgi and Glashow proposed that the molecular states
involving hidden $c\bar{c}$ pair do exist, and have a rich spectrum
\cite{De Rujula:1976qd}. Possible new resonance near the
$D^{*}\bar{D}^{*}$ threshold was suggested by Voloshin
\cite{Voloshin:2006wf,Dubynskiy:2006sg}. Different from other
possible exotic structures, there are uncontroversial evidences for
hadronic molecule such as the deuteron, which is unambiguously a
proton-neutron bound state.  The deuteron has been studied in great
details over the years \cite{Ericson:1985hf,Ericson:1982ei}. From
these studies, we learn that the pion exchange determines most of
the binding energy and the long range part of the deuteron
wavefunction, and the S-D wave mixing effect plays a critical role
in  providing the binding. Guided by the binding of deuteron,
Tornqvist performed a systematic study of possible deuteronlike two
mesons bound states with long distance one pion exchange
\cite{Tornqvist:1991ks,Tornqvist:1993ng}. At short distance, the
interaction should be induced by the interactions among the quarks.
However, a detailed and reliable modelling of the short range
interaction is not a easy matter, and various phenomenological
models have been proposed \cite{Barnes:1991em,Swanson:1992ec},
although one pion exchange is expected to be dominant for the
hadronic molecule. Inspired by the nucleon-nucleon interactions, we
further extended the one pion exchange model to include the short
distance contributions from the heavier bosons $\eta$, $\sigma$,
$\rho$ and $\omega$ exchanges in Ref. \cite{Ding:2009vj}. We have
also taken into account the contribution of the "$\delta$ function"
term, which leads to a $\delta$ function term in the effective
potential in configuration space when no regularization is used.
This one boson exchange model gives a very good description of the
weakly bound hadronic molecule. It has been successfully applied to
exploring the possible heavy flavor pseudoscalar-vector molecular
states \cite{Ding:2009vj}, the $D^{*}_s\bar{D}^{*}_s$ system and the
molecular interpretation of $Y(4140)$ \cite{Ding:2009vd}. Motivated
by the controversial states $Y(4008)$ and $Z^{+}_1(4050)$, we shall
examine for which quantum numbers the boson exchange potential is
attractive and strong enough so that bound states are expected, ten
allowed $D^{*}\bar{D}^{*}$ states with low spin parity are
considered. Moreover, the $B^{*}\bar{B}^{*}$ system would be
discussed as well.

The paper is organized as follows. In section II, the formalism of
one boson exchange model is summarized. In section III, we apply the
one boson exchange model to the $D^{*}\bar{D}^{*}$ system, the
quantum numbers of the $D^{*}\bar{D}^{*}$ bound states which might
exist, are suggested. The $B^{*}\bar{B}^{*}$ system is discussed
along the same line in section IV. Finally we present our
conclusions and some discussions in section V.

\section{The formalism of the one boson exchange model}

In the one boson exchange model, the effective potential between two
hadrons is obtained by summing the interactions between light quarks
or antiquarks via one boson exchange. To leading order in the boson
fields and their derivative, the effective interactions between the
constituent quark and the exchanged boson are as follows
\cite{Nagels:1975fb,Nagels:1977ze,Machleidt:1987hj,Ding:2009vj}
\begin{eqnarray}
\nonumber \rm
{Pseudoscalar:}&&~~~~~\mathcal{L}_p=-g_{pqq}\bar{\psi}(x)i\gamma_5\psi(x)\varphi(x)\\
\nonumber \rm
{Scalar:}&&~~~~~\mathcal{L}_s=-g_{sqq}\bar{\psi}(x)\psi(x)\phi(x)\\
\label{1}\rm{Vector:}&&~~~~~\mathcal{L}_v=-g_{vqq}\bar{\psi}(x)\gamma_{\mu}\psi(x)v^{\mu}(x)-\frac{f_{vqq}}{2m_{q}}\bar{\psi}(x)\sigma_{\mu\nu}\psi(x)\partial^{\mu}v^{\nu}(x)
\end{eqnarray}
Here $m_q$ is the constituent quark mass, $\psi(x)$ is the
constituent quark Dirac spinor field, $\varphi(x)$, $\phi(x)$ and
$v^{\mu}(x)$ are isospin-singlet pseudoscalar, scalar and vector
boson fields respectively. In this work we take $m_q\equiv
m_u=m_d\simeq313$ MeV, since we concentrate on the constituent up
and down quarks here. If the isovector bosons are involved, the
couplings enter in the form $\bm{\tau\cdot\varphi}$,
$\bm{\tau\cdot\phi}$ and $\bm{\tau\cdot v^{\mu}}$ respectively,
where $\bm{\tau}$ is the well-known Pauli matrices. Following
standard procedure, we can straightforwardly obtain the one boson
exchange potential between two quarks.
\begin{enumerate}
\item{Pseudoscalar boson exchange}
\begin{equation}
\label{2}V_p(\mathbf{r})=
\frac{g^2_{pqq}}{4\pi}\frac{\mu^3_p}{12m^2_q}\big[-H_1(\Lambda,m_{p},\mu_p,r)\,\bm{\sigma}_i\cdot\bm{\sigma}_j+H_3(\Lambda,m_{p},\mu_p,r)S_{ij}(\hat{\mathbf{r}})\big]
\end{equation}
where
$S_{ij}(\hat{\mathbf{r}})\equiv3(\bm{\sigma}_i\cdot\hat{\mathbf{r}})(\bm{\sigma}_j\cdot\hat{\mathbf{r}})-\bm{\sigma}_i\cdot\bm{\sigma}_j$
is the tensor operator. We have defined
$\mu^2_p=m^2_p-(m_{V1}-m_{V2})^2$ to approximately account for the
recoil effect due to the small mass splitting within the heavy
flavor vector meson isospin multiplet
\cite{Tornqvist:1991ks,Tornqvist:1993ng,Thomas:2008ja}, where $m_p$
is the exchanged pseudoscalar mass, $m_{V1}$ and $m_{V2}$ are the
masses of the heavy flavor vector mesons involved. For the
$B^{*}\bar{B}^{*}$ system, the mass splitting can be negligible so
that $\mu_p\simeq m_p$ is satisfied.
\item{Scalar boson exchange}
\begin{equation}
\label{3}V_s(\mathbf{r})=-\mu_s\frac{g^2_{sqq}}{4\pi}\left[H_0(\Lambda,m_s,\mu_s,r)+\frac{\mu^2_s}{8m^2_q}H_1(\Lambda,m_s,\mu_s,r)+\frac{\mu^2_s}{2m^2_q}H_2(\Lambda,m_s,\mu_s,r)\mathbf{L}\cdot\mathbf{S}_{ij}\right]
\end{equation}
where $\mathbf{S}_{ij}=\frac{1}{2}(\bm{\sigma}_i+\bm{\sigma}_j)$,
$\mu^2_s=m^2_s-(m_{V1}-m_{V2})^2$ with $m_s$ being the exchanged
scalar meson mass, and $\mathbf{L}=\mathbf{r}\times\mathbf{p}$ is
the angular momentum operator.
\item{Vector boson exchange}
\begin{eqnarray}
\nonumber
V_v(\mathbf{r})&=&\frac{\mu_v}{4\pi}\bigg\{g^2_{vqq}H_0(\Lambda,m_v,\mu_v,r)-(g^2_{vqq}+4g_{vqq}f_{vqq})\frac{\mu^2_v}{8m^2_q}H_1(\Lambda,m_v,\mu_v,r)\\
\nonumber&&-(g_{vqq}+f_{vqq})^2\frac{\mu^2_v}{12m^2_q}\Big[H_3(\Lambda,m_v,\mu_v,r)S_{ij}(\hat{\mathbf{r}})+2H_1(\Lambda,m_v,\mu_v,r)(\bm{\sigma_i\cdot\bm{\sigma}_j})\Big]\\
\label{4}&&-(3g^2_{vqq}+4g_{vqq}f_{vqq})\frac{\mu^2_{v}}{2m^2_q}H_2(\Lambda,m_v,\mu_v,r)\mathbf{L}\cdot\mathbf{S}_{ij}\bigg\}
\end{eqnarray}
where $\mu^2_v=m^2_v-(m_{V1}-m_{V2})^2$ approximately reflects the
recoil effect with $m_v$ being the exchanged vector meson mass. For
$I=1$ isovector boson exchange, the above three potentials in
Eq.(\ref{2})-Eq.(\ref{4}) should be multiplied by the operator
$\bm{\tau}_i\cdot\bm{\tau}_j$ in the isospin space.
\end{enumerate}
The dimensionless functions $H_0(\Lambda,m,\mu,r)$,
$H_1(\Lambda,m,\mu,r)$, $H_2(\Lambda,m,\mu,r)$ and
$H_3(\Lambda,m,\mu,r)$ introduced in Eq.(\ref{2})-Eq.(\ref{4}) are
defined as follows
\begin{eqnarray}
\nonumber H_0(\Lambda,m,\mu,r)&=&\frac{1}{\mu r}\big(e^{-\mu
r}-e^{-Xr}\big)-\frac{\Lambda^2-m^2}{2\mu X}\,e^{-Xr}\\
\nonumber H_1(\Lambda,m,\mu,r)&=&-\frac{1}{\mu r}\big(e^{-\mu
r}-e^{-Xr}\big)+\frac{X(\Lambda^2-m^2)}{2\mu^3}\,e^{-Xr}\\
\nonumber H_2(\Lambda,m,\mu,r)&=&\big(1+\frac{1}{\mu
r}\big)\frac{1}{\mu^2r^2}e^{-\mu
r}-\big(1+\frac{1}{Xr}\big)\frac{X}{\mu}\frac{1}{\mu^2r^2}e^{-Xr}-\frac{\Lambda^2-m^2}{2\mu^2}\frac{e^{-Xr}}{\mu
r}\\
\label{5} H_3(\Lambda,m,\mu,r)&=&\big(1+\frac{3}{\mu
r}+\frac{3}{\mu^2r^2}\big)\frac{1}{\mu r}e^{-\mu
r}-\big(1+\frac{3}{Xr}+\frac{3}{X^2r^2}\big)\frac{X^2}{\mu^2}\frac{e^{-Xr}}{\mu
r}-\frac{\Lambda^2-m^2}{2\mu^2}\big(1+Xr\big)\frac{e^{-Xr}}{\mu r}
\end{eqnarray}
with $X^2=\Lambda^2+\mu^2-m^2$. In deriving the above effective
potentials, we have introduced form factor at each interaction
vertex to regularize the effective potential at short distance, and
the form factor in momentum space is taken as
\begin{equation}
\label{6}F(q)=\frac{\Lambda^2-m^2}{\Lambda^2-q^2}
\end{equation}
where $\Lambda$ is the so-called regularization parameter, $m$ and
$q$ are the mass and the four momentum of the exchanged boson
respectively. This form factor suppresses the contribution of high
momentum, i.e. small distance. The presence of such a form factor is
dictated by the extended structure of the hadrons. The parameter
$\Lambda$, which governs the range of suppression, can be directly
related to the hadron size which is approximately proportional to
$1/\Lambda$. However, since the question of hadron size is still
very much open, the value of $\Lambda$ is poorly known
phenomenologically, and it is dependent on the models and
applications. In the nucleon-nucleon interactions, the $\Lambda$ in
the range of 0.8-1.5 GeV has been used to fit the data. For the
present application to the heavy flavor vector mesons system, which
have a smaller size than the nucleon, we would expect a larger
regularization parameter $\Lambda$. We have demonstrated that the
binding energy and static properties of the deuteron are produced
very well in the one boson exchange model, if $\Lambda$ is chosen to
be about 808 MeV \cite{Ding:2009vj}. In the case that all coupling
constants except $g_{\pi NN}$ are reduced by half, $\Lambda$ should
be approximately $970$ MeV. The extended structure of hadrons also
has the following obvious consequence: because the mass of the
exchanged meson determines the range of the corresponding
contribution to the $D^{*}\bar{D}^{*}$ interactions, one should
restrict oneself to meson exchange with the exchanged meson mass
below a certain value, typically on the order of the regularization
parameter $\Lambda$. Since $\pi$, $\eta$, $\sigma$, $\rho$ and
$\omega$ exchanges are considered in the present work, the value of
$\Lambda$ should be larger than the $\omega$ meson mass.

In the present one boson exchange model, the input parameters
include the masses of the exchanged bosons and heavy flavor vector
mesons, and the effective coupling constants between the constituent
quarks and the exchanged bosons. The meson masses are taken from the
compilation of the Particle Data Group \cite{pdg}:
$m_{\pi^{\pm}}=139.57$ MeV, $m_{\pi^{0}}=134.98$ MeV,
$m_{\eta}=547.85$ MeV, $m_{\sigma}=600$ MeV, $m_{\rho}=775.49$ MeV,
$m_{\omega}=782.65$ MeV, ${m_{D^{*0}}=2006.97}$ MeV,
${m_{D^{*\pm}}=2010.27}$ MeV and $m_{B^{*}}=5325.1$ MeV. The
constituent quark-meson coupling constants can be estimated from the
phenomenologically known $\pi NN$, $\eta NN$, $\sigma NN$, $\rho NN$
and $\omega NN$ coupling constants via the well-known
Goldberger-Treiman relation \cite{Ding:2009vj,Riska:2000gd}.
\begin{eqnarray}
\nonumber&&g_{\pi qq}=\frac{3}{5} \frac{m_q}{m_N}\,g_{\pi
NN},~~~~g_{\eta qq}=\frac{m_q}{m_{N}}\,g_{\eta NN}\\
\nonumber&&g_{\rho qq}=g_{\rho NN},~~~~~~~~~~~~f_{\rho
qq}=\frac{3}{5}\frac{m_q}{m_{N}}f_{\rho
NN}-(1-\frac{3}{5}\frac{m_q}{m_N})g_{\rho NN}\\
\nonumber&&g_{\omega qq}=\frac{1}{3}\,g_{\omega
NN},~~~~~~~~f_{\omega qq}=\frac{m_q}{m_N}f_{\omega
NN}-(\frac{1}{3}-\frac{m_q}{m_N})g_{\omega NN}\\
\label{7}&&g_{\sigma qq}=\frac{1}{3}\,g_{\sigma NN}
\end{eqnarray}
Among the above effective boson-nucleon coupling constants, only
$g_{\pi NN}$ has been determined accurately from the pion-nucleon
and nucleon-nucleon scatterings. As in our previous work
\cite{Ding:2009vj,Ding:2009vd}, the effective coupling constants are
taken from the famous Bonn model \cite{Machleidt:1987hj}. The
uncertainty of the coupling constants will be taken into account
later, all the coupling constants except $g_{\pi NN}$ are reduced by
half for demonstration, and the corresponding numerical results are
presented as well.

For the system consisting of two vector mesons, the spatial parity
is determined by $P=(-1)^{L}$ and the $C$ parity is $C=(-1)^{L+S}$,
where $L$ is the relative angular momentum between two vector
mesons, and $S$ is the total spin of the system. We cutoff the total
angular momentum of the system to $J=2$, the allowed states with low
spin parity are listed in Table \ref{states}. In the following, we
shall study for which quantum numbers the one boson exchange
potential is so attractive that bound states may exist.

\begin{center}
\begin{table}[hptb]
\begin{tabular}{|cl|}\hline\hline
$J^{PC}$&~~~~~Channels\\\hline
$0^{++}$    & ~~~~~$^1S_0$, $^5D_0$    \\
$1^{+-}$    & ~~~~~$^3S_1$,$^3D_1$   \\
$0^{-+}$    &~~~~~ $^3P_0$   \\
$1^{++}$    &~~~~~ $^5D_1$     \\
$1^{-+}$    &~~~~~ $^3P_1$    \\
$2^{+-}$    &~~~~~ $^3D_2$    \\
$1^{--}$    &~~~~~ $^1P_1$, $^5P_1$, $^5F_1$      \\
$2^{++}$    &~~~~~ $^1D_2$, $^5S_2$, $^5D_2$, $^5G_2$        \\
$2^{-+}$    & ~~~~~$^3P_2$, $^3F_2$      \\
$2^{--}$    &~~~~~ $^5P_2$, $^5F_2$      \\
\hline \hline
\end{tabular}
\caption{\label{states}The allowed states of the system consisting
of two vector mesons, where only states whose total angular momentum
is smaller than 3 are listed.}
\end{table}
\end{center}

\section{The possible molecular states of the $D^{*}\bar{D}^{*}$ system}

\begin{figure}
\includegraphics[scale=.645]{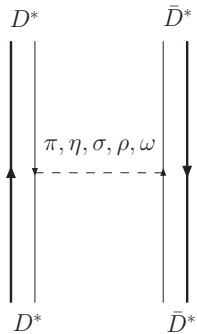}
\caption{$D^{*}\bar{D}^{*}$ interactions in the one boson exchange
model at quark level, where the thick line represents heavy quark or
antiquark, and the thin line denotes light quark or antiquark.}
\label{interaction diagram}
\end{figure}
\begin{figure}[hptb]
\includegraphics[scale=.745]{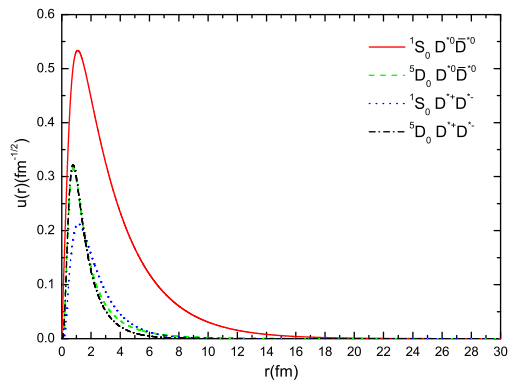}
\caption{\label{charm_0++} The four components spatial wavefunctions
of the $0^{++}$ $D^{*}\bar{D}^{*}$ state with $\Lambda=940$ MeV. }
\end{figure}

There is a sign difference $(-1)^{G}$ between the quark-quark
interactions and quark-antiquark interactions, the magnitudes are
the same, where $G$ is the $G$-parity of the exchanged meson.
Consequently both $\pi$ and $\omega$ exchanges give opposite sign
between quark-quark interactions and quark-antiquark interactions.
In the present case, the effective potential is induced by one boson
exchange between a pair of light quark and antiquark, and the
diagram contributing to $D^{*}\bar{D}^{*}$ interactions is shown in
Fig. \ref{interaction diagram}. The effective potential is
explicitly expressed as
\begin{eqnarray}
\nonumber V(\mathbf{r})&=&-V_{\pi}(\mathbf{r})+V_{\eta}(\mathbf{r})+V_{\sigma}(\mathbf{r})+V_{\rho}(\mathbf{r})-V_{\omega}(\mathbf{r})\\
\nonumber&\equiv&
V_C(r)+V_S(r)(\bm{\sigma}_i\cdot\bm{\sigma}_j)+V_I(\mu_{\rho},r)(\bm{\tau}_i\cdot\bm{\tau}_j)+V_T(r)S_{ij}(\mathbf{\hat{r}})+
V_{SI}(m_{\pi},\mu_{\pi},\mu_{\rho},r)(\bm{\sigma}_i\cdot\bm{\sigma}_j)(\bm{\tau}_i\cdot\bm{\tau}_j)\\
\label{8}&&+V_{TI}(m_{\pi},\mu_{\pi},\mu_{\rho},r)S_{ij}(\mathbf{\hat{r}})(\bm{\tau}_i\cdot\bm{\tau}_j)+V_{LS}(r)(\mathbf{L}\cdot\mathbf{S}_{ij})+V_{LSI}
(\mu_{\rho},r)(\mathbf{L}\cdot\mathbf{S}_{ij})(\bm{\tau}_i\cdot\bm{\tau}_j)
\end{eqnarray}
where $V_{\cal M}(r)$ (${\cal M}=\pi$, $\eta$, $\sigma$, $\rho$ and
$\omega$) denotes the effective potential induced by the meson
${\cal M}$ exchange between two quarks. The subscripts $i$ and $j$
are the indexes of light quark and antiquark. The spin
operator(isospin operator) $\bm{\sigma}_i$ or $\bm{\sigma}_j$
($\bm{\tau}_i$ or $\bm{\tau}_j$) only acts on the light quark and
antiquark. The parameters $m_{\pi}$, $\mu_{\pi}$ and $\mu_{\rho}$
could take two different sets of values due to the small mass
splitting within the $D^{*}$ and $\pi$ isospin multiplets. For
$D^{*0}\bar{D}^{*0}\rightarrow D^{*0}\bar{D}^{*0}$ and
$D^{*+}D^{*-}\rightarrow D^{*+}D^{*-}$, we should choose
$m_{\pi}=m_{\pi^{0}}$, $\mu_{\pi}=m_{\pi^{0}}$ and
$\mu_{\rho}=m_{\rho}$. Whereas for the
$D^{*0}\bar{D}^{*0}\rightarrow D^{*+}D^{*-}$ and
$D^{*+}D^{*-}\rightarrow D^{*0}\bar{D}^{*0}$ processes, we should
take $m_{\pi}=m_{\pi^{\pm}}$,
$\mu_{\pi}=[m^2_{\pi^{\pm}}-(m_{D^{*+}}-m_{D^{*0}})^2]^{1/2}\equiv\mu_{\pi_1}$
and
$\mu_{\rho}=[m^2_{\rho}-(m_{D^{*+}}-m_{D^{*0}})^2]^{1/2}\equiv\mu_{\rho_1}$.
The eight potential functions $V_C(r)$, $V_S(r)$ etc are given by
\begin{eqnarray}
\nonumber &&V_C({r})=-\frac{g^2_{\sigma
qq}}{4\pi}m_{\sigma}\Big[H_0(\Lambda,m_{\sigma},m_{\sigma},r)+\frac{m^2_{\sigma}}{8m^2_q}H_1(\Lambda,m_{\sigma},m_{\sigma},r)\Big]-\frac{g^2_{\omega
qq}}{4\pi}m_{\omega}H_0(\Lambda,m_{\omega},m_{\omega},r)\\
\nonumber&&+\frac{g^2_{\omega qq}+4g_{\omega qq}f_{\omega
qq}}{4\pi}\frac{m^3_{\omega}}{8m^2_q}H_1(\Lambda,m_{\omega},m_{\omega},r)\\
\nonumber &&V_S({r})=-\frac{g^2_{\eta
qq}}{4\pi}\frac{m_{\eta}^3}{12m^2_q}H_1(\Lambda,m_{\eta},m_{\eta},r)+\frac{(g_{\omega
qq}+f_{\omega
qq})^2}{4\pi}\frac{m_{\omega}^3}{6m^2_q}H_1(\Lambda,m_{\omega},m_{\omega},r)\\
\nonumber && V_I(\mu_{\rho},{r})=\frac{g^2_{\rho
qq}}{4\pi}\,\mu_{\rho}H_0(\Lambda,m_{\rho},\mu_{\rho},r)-\frac{g^2_{\rho
qq}+4g_{\rho qq}f_{\rho qq}}{4\pi}\frac{\mu^3_{\rho}}{8m^2_q}H_1(\Lambda,m_{\rho},\mu_{\rho},r)\\
\nonumber &&V_{T}({r})=\frac{g^2_{\eta
qq}}{4\pi}\frac{m_{\eta}^3}{12m^2_q}H_3(\Lambda,m_{\eta},m_{\eta},r)+\frac{(g_{\omega
qq}+f_{\omega
qq})^2}{4\pi}\frac{m_{\omega}^3}{12m^2_q}H_3(\Lambda,m_{\omega},m_{\omega},r)\\
\nonumber &&
V_{SI}(m_{{\pi}},\mu_{\pi},\mu_{{\rho}},{r})=\frac{g^2_{\pi
qq}}{4\pi}\frac{\mu^3_{\pi}}{12m^2_{q}}H_1(\Lambda,m_{\pi},\mu_{\pi},r)-\frac{(g_{\rho
qq}+f_{\rho
qq})^2}{4\pi}\frac{\mu^3_{\rho}}{6m^2_q}H_1(\Lambda,m_{\rho},\mu_{\rho},r)\\
\nonumber &&V_{TI}(m_{\pi},\mu_{\pi},\mu_{\rho},r)= -\frac{g^2_{\pi
qq}}{4\pi}\frac{\mu^3_{\pi}}{12m^2_q}H_3(\Lambda,m_{\pi},\mu_{\pi},r)-\frac{(g_{\rho
qq}+f_{\rho
qq})^2}{4\pi}\frac{\mu^3_{\rho}}{12m^2_q}H_3(\Lambda,m_{\rho},\mu_{\rho},r)\\
\nonumber && V_{LS}(r)=-\frac{g^2_{\sigma
qq}}{4\pi}\frac{m^3_{\sigma}}{2m^2_q}H_2(\Lambda,m_{\sigma},m_{\sigma},r)+\frac{3g^2_{\omega
qq}+4g_{\omega qq}f_{\omega
qq}}{4\pi}\frac{m^3_{\omega}}{2m^2_q}H_2(\Lambda,m_{\omega},m_{\omega},r)\\
\label{9}&& V_{LSI}(\mu_{\rho},r)=-\frac{3g^2_{\rho qq}+4g_{\rho
qq}f_{\rho
qq}}{4\pi}\frac{\mu^3_{\rho}}{2m^2_q}H_2(\Lambda,m_{\rho},\mu_{\rho},r)
\end{eqnarray}
Since the threshold of $D^{*+}D^{*-}$ is about 6.6 MeV higher than
the $D^{*0}\bar{D}^{*0}$ threshold, the isospin symmetry is expected
to be violated drastically for the $D^{*}\bar{D}^{*}$ molecular
states whose binding energy is of order a few MeV
\cite{Ding:2009vj}. Isospin violation mainly comes from three
aspects: the first is the different kinetic energies for
$D^{*0}\bar{D}^{*0}$ and $D^{*+}D^{*-}$, the second is the different
effective potentials from $\pi^{0}$ exchange and $\pi^{\pm}$
exchange, and the third is because of the different thresholds of
$D^{*0}\bar{D}^{*0}$ and $D^{*+}D^{*-}$. In the following, we will
perform the same analysis as that for the deuteron and the possible
heavy flavor molecules in Ref. \cite{Ding:2009vj,Ding:2009vd}. One
can then determine for which quantum numbers the one boson exchange
potential is attractive and strong enough so that the
$D^{*}\bar{D}^{*}$ bound states are expected. Firstly we consider
the $J^{PC}=0^{++}$ $D^{*}\bar{D}^{*}$ states as an demonstration,
the system can be in S wave or D wave similar to the deuteron.
Taking into account the isospin violation effect, one has four
coupled channels. For convenience, we choose the basis to be
$|1\rangle\equiv|^1S_0(D^{*0}\bar{D}^{*0})\rangle$,
$|2\rangle\equiv|^5D_0(D^{*0}\bar{D}^{*0})\rangle$,
$|3\rangle\equiv|^1S_0(D^{*+}D^{*-})\rangle$ and
$|4\rangle\equiv|^5D_0(D^{*+}D^{*-})\rangle$, then the wavefunction
of the system is written as
\begin{eqnarray}
\label{10}
|0^{++}(D^{*}\bar{D}^{*})\rangle=\frac{u_1(r)}{r}|^1S_0(D^{*0}\bar{D}^{*0})\rangle+\frac{u_2(r)}{r}|^5D_0(D^{*0}\bar{D}^{*0})\rangle+\frac{u_3(r)}{r}|^1S_0(D^{*+}D^{*-})\rangle+\frac{u_4(r)}{r}|^5D_0(D^{*+}D^{*-})\rangle
\end{eqnarray}
where $u_1(r)$, $u_2(r)$, $u_3(r)$ and $u_4(r)$ are the spatial
wavefunctions. The matrix elements of the light quark relevant
operators $\bm{\sigma}_i\cdot\bm{\sigma}_j$,
$S_{ij}(\hat{\mathbf{r}})$ and $\mathbf{L}\cdot\mathbf{S}_{ij}$ etc
in Eq.(\ref{8}) can be calculated straightforwardly with the help of
angular momentum algebra, and the results are given analytically in
the Appendix of Ref.\cite{Ding:2009vj}. Consequently the one boson
exchange potential for the $0^{++}$ $D^{*}\bar{D}^{*}$ state can be
written in the matrix form as
\begin{eqnarray}
\nonumber&& V_{0^{++}}(r)=V_{C}(r)\left(\begin{array}{cccc}
1&0&0&0\\
0&1&0&0\\
0&0&1&0\\
0&0&0&1
\end{array} \right)+V_{S}(r)\left(\begin{array}{cccc}
-2&0&0&0\\
0&1&0&0\\
0&0&-2&0\\
0&0&0&1
\end{array} \right)+V_{I}(\mu_{\rho},r)\left(\begin{array}{cccc}
-1&0&-2&0\\
0&-1&0&-2\\
-2&0&-1&0\\
0&-2&0&-1
\end{array} \right)\\
\nonumber&&+V_{T}(r)\left(\begin{array}{cccc}
0&-\sqrt{2}&0&0\\
-\sqrt{2}&-2&0&0\\
0&0&0&-\sqrt{2}\\
0&0&-\sqrt{2}&-2
\end{array} \right)+V_{SI}(m_{\pi},\mu_{\pi},\mu_{\rho},r)\left(\begin{array}{cccc}
2&0&4&0\\
0&-1&0&-2\\
4&0&2&0\\
0&-2&0&-1
\end{array} \right)\\
\label{10}&&+V_{TI}(m_{\pi},\mu_{\pi},\mu_{\rho},r)\left(\begin{array}{cccc}
0&\sqrt{2}&0&2\sqrt{2}\\
\sqrt{2}&2&2\sqrt{2}&4\\
0&2\sqrt{2}&0&\sqrt{2}\\
2\sqrt{2}&4&\sqrt{2}&2
\end{array} \right)+V_{LS}(r)\left(\begin{array}{cccc}
0&0&0&0\\
0&-3&0&0\\
0&0&0&0\\
0&0&0&-3
\end{array} \right)+V_{LSI}(\mu_{\rho},r)\left(\begin{array}{cccc}
0&0&0&0\\
0&3&0&6\\
0&0&0&0\\
0&6&0&3
\end{array} \right)
\end{eqnarray}
For the up-left and down-right $2\times2$ matrix elements, we should
choose $m_{\pi}=m_{\pi^{0}}$, $\mu_{\pi}=m_{\pi^{0}}$ and
$\mu_{\rho}=m_{\rho}$. While for the off-diagonal $2\times2$ matrix
elements, we should take $m_{\pi}=m_{\pi^{\pm}}$,
$\mu_{\pi}=\mu_{\pi_1}$ and $\mu_{\rho}=\mu_{\rho_1}$. Taking into
account the D wave centrifugal barrier and solving the coupled
channel Schr$\ddot{\rm o}$dinger equation numerically, the numerical
results are listed in the Table \ref{charm0++}. It is obvious that
the binding energy and the static properties are rather sensitive to
the regularization parameter $\Lambda$ and the effective coupling
constants, this is common to the one boson exchange model
\cite{Ding:2009vj,Ding:2009vd, Thomas:2008ja}. We also find that the
binding energy increases with $\Lambda$, this is because increasing
$\Lambda$ increases the strength of the potential at short distance.
For $\Lambda=930$ MeV, a bound state with mass about 4013.80 MeV
appears. We can see that the isospin symmetry is strongly broken
especially for the states near the $D^{*}\bar{D}^{*}$ threshold.
Fig. \ref{charm_0++} displays the wavefunction of the bound state
with mass 4004.40 MeV and $\Lambda=950$ MeV. One notice that the
$D^{*0}\bar{D}^{*0}$ component dominates over the $D^{*+}D^{*-}$
component for both the S wave and D wave configurations, as could be
expected. Because the wavefunctions $u_1(r)$ and $u_3(r)$ have the
same sign, the same is true for $u_2(r)$ and $u_4(r)$. Therefore the
$I=0$ component is predominant for this state, it would be an
isospin singlet in the isospin symmetry limit. The dominance of the
$I=0$ configuration is observed for all the states listed in Table
\ref{charm0++}. From the numerical results, we can see that the D
wave probability increases with the regularization parameter
$\Lambda$, the importance of the tensor force is obvious. The
uncertainties induced by the effective coupling constants are
considered as well. All the coupling constants except $g_{\pi NN}$
are reduced by half, and the corresponding numerical results are
presented in Table \ref{charm0++}. The same pattern of the static
properties dependence on $\Lambda$ is found. Bound state solution
appears if the regularization parameter $\Lambda$ is about $1100$
MeV, the value of $\Lambda$ is still in the reasonable range. Since
the molecular state is widely extended, the decay into light mesons
via annihilation is generally suppressed by the form factor. The
leading source of decay is dissociation, to a good approximation the
dissociation will proceed via the almost free space decay of the
constituent mesons. Consequently the $0^{++}$ $D^{*}\bar{D}^{*}$
molecule mainly decays into $D\bar{D}\gamma\gamma$, and
$D\bar{D}\gamma\pi$, and the mode $D\bar{D}\pi\pi$ is strongly
suppressed or even forbidden by the phase space.

The one boson exchange potentials for the nine remaining states are
listed in Appendix A. Following exactly the same method, the binding
energy and the static properties can be predicted, and partial
results are shown in Appendix B for illustration. The binding energy
and the static properties are found to be rather sensitive the
regularization parameter $\Lambda$. The spatial wavefunctions for
the $D^{*0}\bar{D}^{*0}$ and $D^{*+}D^{*-}$ components have the same
sign, consequently the $I=0$ component is dominant for all these
states, and they are isospin singlets in the isospin symmetry limit.
The same conclusion has been reached in the one pion exchange model
\cite{Tornqvist:1993ng}, where the strength of the effective
potential for the $I=0$ state is about one third of that for $I=1$.
We shall discuss these states one by one in the following.

For the axial vector $1^{+-}$ state, there are four channels
$^3S_1$($D^{*0}\bar{D}^{*0}$), $^3D_1$($D^{*0}\bar{D}^{*0}$),
$^3S_1$($D^{*+}D^{*-}$) and $^3D_1$($D^{*+}D^{*-}$). The energy of
the system is substantially lowered due to the S-D wave mixing
effect. The coupling between the S wave and D wave has the same
strength as the $0^{++}$ case, which is clearly seen from
Eq.(\ref{10}) and Eq.(\ref{a1}). It is obvious that the predictions
for the binding energy and the static properties have similar
pattern with the ones for the $0^{++}$ state, and the binding of the
$1^{+-}$ state is less stronger than the $0^{++}$ one for the same
$\Lambda$ value. For $\Lambda$ as large as 980 MeV, we can find a
bound state with mass about 4012.43 MeV. If all coupling constants
except $g_{\pi NN}$ are reduced by half, bound state begins to
appear for $\Lambda\simeq1200$ MeV. We note that the unnatural spin
parity forbids its decay into $D\bar{D}$, while the decay mode
$D\bar{D}^{*}/D^{*}\bar{D}$ is allowed.

The pseudoscalar $0^{-+}$ $D^{*}\bar{D}^{*}$ state involve two
channels $^3P_0$($D^{*0}\bar{D}^{*0}$) and $^3P_0$($D^{*+}D^{*-}$).
If the small isospin violation effect is neglected, the two coupled
channel problem is reduced to a familiar one channel problem.
Although there is repulsive P wave centrifugal barrier, the one
boson exchange potential is so strong that the P wave centrifugal
barrier can be partly compensated, therefore bound state solutions
can be found for reasonable values of $\Lambda$, as can be seen from
Table \ref{charm0-+}. For $\Lambda=950-1030$ MeV, we find that the
binding energy with respect to the $D^{*0}\bar{D}^{*0}$ threshold is
in the range of 2.6 to 137.2 MeV. The binding energy is more
sensitive to $\Lambda$ than the $0^{++}$ and $1^{+-}$ four coupled
channels cases. The probabilities for the
$^3P_0$($D^{*0}\bar{D}^{*0}$) and $^3P_0$($D^{*+}D^{*-}$) components
are close to each other, as could be expected. The small difference
is induced by the mass splitting within the $D^{*}$ and $\pi$
isospin multiplet.

The results for the $1^{++}$ state are similar to the $0^{-+}$ case.
Because the D wave centrifugal barrier is higher than the P wave
centrifugal barrier, the total effective potential for the $1^{++}$
state is less attractive than the $0^{-+}$ state. If the coupling
constants except $g_{\pi NN}$ are reduced by half, bound state
solutions can be found only for $\Lambda$ larger than 1290 MeV.
Since the numerical results indicate that larger value of $\Lambda$
is required to bind the $1^{++}$ $D^{*}\bar{D}^{*}$ state, the
$1^{++}$ $D^{*}\bar{D}^{*}$ state is harder to be bound than the
previous states considered.

For the CP exotic $1^{-+}$ and $2^{+-}$ states, the system is in P
wave and D wave respectively. Considering isospin violation effect,
two channels are involved for both states. As has been shown in
Eq.(\ref{a4}) and Eq.(\ref{a5}), the one boson exchange potentials
for these two states are exactly the same, and they are less
attractive than the potentials for the $0^{-+}$ and $1^{++}$ states.
For the $1^{-+}$ state, bound state solution appears only for
$\Lambda$ as large as 1300 MeV, and we can find $2^{+-}$ bound state
only if the regularization parameter $\Lambda$ is larger than 1640
MeV. If we reduce all the coupling constants except $g_{\pi NN}$ by
half, $\Lambda$ larger than 3010 MeV and 5110 MeV respectively for
the $1^{-+}$ and $2^{+-}$ states is required to find bound state
solutions. Because the value of $\Lambda$ is so large that it is far
beyond the range of 0.8 to 1.5 GeV favored by the nucleon-nucleon
interactions, we tend to conclude that the CP exotic $1^{-+}$ and
$2^{+-}$ $D^{*}\bar{D}^{*}$ states can not be bound by the one boson
exchange potential. This conclusion is consistent with the fact that
no such CP exotic states have been observed so far.

\begin{figure}[hptb]
\includegraphics[scale=.745]{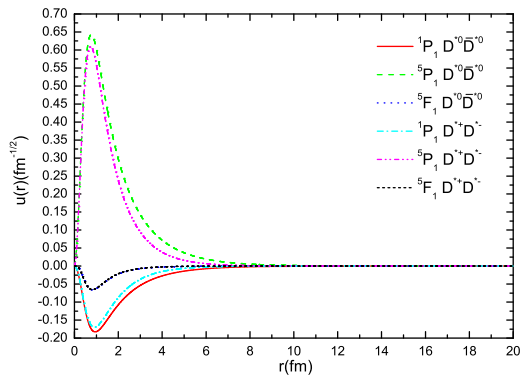}
\caption{\label{charm_1--} The spatial wavefunctions of the $1^{--}$
$D^{*}\bar{D}^{*}$ state with $\Lambda=920$ MeV. }
\end{figure}

We then come to the very interesting $1^{--}$ $D^{*}\bar{D}^{*}$
state, there are six configurations $^1P_1$($D^{*0}\bar{D}^{*0}$),
$^5P_1$($D^{*0}\bar{D}^{*0}$), $^5F_1$($D^{*0}\bar{D}^{*0}$),
$^1P_1$($D^{*+}D^{*-}$), $^5P_1$($D^{*+}D^{*-}$) and $^5F_1$(
$D^{*+}D^{*-}$). Due to the substantially strong attraction of the
effective potential, bound state solutions can be found for
reasonable value of $\Lambda$ in spite of the P wave centrifugal
barrier. The wavefunction of the bound state with mass about 4006.82
MeV and $\Lambda=920$ MeV is displayed in Fig. \ref{charm_1--}. From
the numerical results in Table \ref{charm1--}, we see that the
isospin symmetry is violated, especially for the states near the
threshold. The $^5P_1$($D^{*0}\bar{D}^{*0}$) and $^5P_1$
($D^{*+}D^{*-}$) are the dominant components, the $^5F_1$ components
are strongly suppressed by the large F wave centrifugal barrier. For
the same value of the regularization parameter $\Lambda$, we notice
that the binding energy of $1^{--}$ state is the largest among the
ten allowed $D^{*}\bar{D}^{*}$ states. Hence we suggest that the
$1^{--}$ $D^{*}\bar{D}^{*}$ molecular state should exist, this
conclusion is consistent with the results obtained from the general
quantum mechanical properties of unitarity and analyticity
\cite{Voloshin:2006wf,Dubynskiy:2006sg}. The $1^{--}$
$D^{*}\bar{D}^{*}$ state is remarkable, it can be directly produced
via the $e^{+}e^{-}$ annihilation or with the help of the initial
state radiation (ISR) technique at $B$ factory. The existence of
such a state can be either confirmed or rejected if more detailed
$e^{+}e^{-}$ annihilation data near the $D^{*}\bar{D}^{*}$ threshold
become available. We strongly urge the Babar and Belle collaboration
to search for this state. The $1^{--}$ $D^{*}\bar{D}^{*}$ molecule
mainly decays into $D\bar{D}\gamma\gamma$ and $D\bar{D}\gamma\pi$
via the dissociation of $D^{*}$ and $\bar{D}^{*}$, the decays into
$D\bar{D}$ and $D\bar{D}^{*}/D^{*}\bar{D}$ are allowed as well. The
width of the $1^{--}$ $D^{*}\bar{D}^{*}$ molecule should be of the
same order as the $D^{*}$ width. Therefore it would be be a narrow
state, and its width is expected to be of the order about 10 MeV.
For the $1^{--}$ state $Y(4008)$ reported by the Belle
collaboration, although it is close to the $D^{*}\bar{D}^{*}$
threshold, its width is huge, which is $\Gamma=(226\pm44\pm87)$ MeV.
Consequently it seems unreasonable to interpret $Y(4008)$ as the
$1^{--}$ $D^{*}\bar{D}^{*}$ molecule.

\begin{figure}[hptb]
\includegraphics[scale=.745]{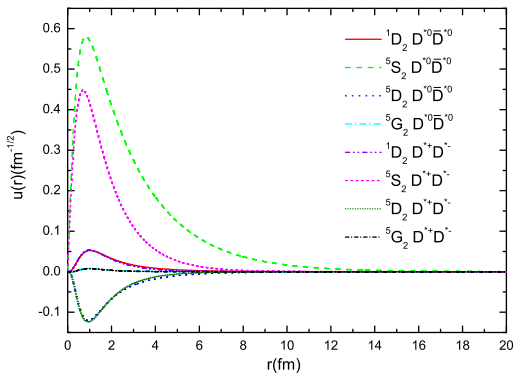}
\caption{\label{charm_2++} The spatial wavefunctions of the $2^{++}$
$D^{*}\bar{D}^{*}$ state with mass 4010.82 MeV and $\Lambda=860$
MeV. }
\end{figure}

For the $2^{++}$ state, eight channels
$^1D_2$($D^{*0}\bar{D}^{*0}$), $^5S_2$($D^{*0}\bar{D}^{*0}$),
$^5D_2$($D^{*0}\bar{D}^{*0}$), $^5G_2$($D^{*0}\bar{D}^{*0}$),
$^1D_2$($D^{*+}D^{*-}$), $^5S_2$($D^{*+}D^{*-}$), $^5D_2$(
$D^{*+}D^{*-}$) and $^5G_2$($D^{*+}D^{*-}$) are involved. The
effective potential in matrix form is given in Eq.(\ref{a7}), which
is more complex than the previous cases considered. It is obvious
that both the tensor interaction and the spin-orbit interaction
vanish in the $^5S_2$ configuration. However, bound state solution
can be found for reasonable value of $\Lambda$ ($\Lambda$ should be
larger than 860 MeV and 970 MeV respectively for the two sets of
coupling constant values). The reason is that the mixing of $^5S_2$
with $^1D_2$, $^5D_2$ and $^5G_2$ under the tensor force increases
the binding of the system considerably through higher order
iterative processes. We show the wavefunction of the $2^{++}$
molecular state with mass 4010.82 MeV and $\Lambda=860$ MeV in Fig.
\ref{charm_2++}. Isospin violation is obvious, and the $I=0$
component is dominant. From the numerical results presented in Table
\ref{charm2++}, we can see that $^5S_2$($D^{*0}\bar{D}^{*0}$) and
$^5S_2$($D^{*+}D^{*-}$) are the dominant components, the $^5D_2$
probability is larger than the $^1D_2$ probability, and the $^5G_2$
component is suppressed. It is remarkable that the dependence of the
$2^{++}$ state mass on $\Lambda$ is the least sensitive among the
ten allowed states. Comparing the numerical results presented in
Table \ref{charm0++} and Table \ref{charm2++}, it can be seen that
the binding of this state is comparable to the $0^{++}$ state,
accordingly the $2^{++}$ $D^{*}\bar{D}^{*}$ molecule very likely
exist.

Finally two states with $J^{PC}=2^{-+}$ and $2^{--}$ respectively
remain. For both states, one has four coupled channels, P wave and F
wave configurations are involved. Because of the P wave and F wave
centrifugal barrier, bound state begins to appear for $\Lambda$ as
large as 1150 MeV and 1090 MeV respectively. If all the coupling
constants except $g_{\pi NN}$ are reduced by half, the value of
$\Lambda$ should be larger than 1630 MeV and 1480 MeV respectively
in order to find bound state solutions. We tend to conclude that the
one boson exchange potential may not support the $2^{-+}$ and
$2^{--}$ $D^{*}\bar{D}^{*}$ states.

In short summary, ten allowed $D^{*}\bar{D}^{*}$ states with low
spin parity has been studied. We find that the isospin symmetry is
violated, especially for the states near the $D^{*}\bar{D}^{*}$
threshold. The $I=0$ configuration is dominant for all the lowest
bound states, hence they would be isospin singlets in the isospin
symmetry limit. Since $Z^{+}_1(4050)$ is an isospin vector,
$Z^{+}_1(4050)$ as a $D^{*}\bar{D}^{*}$ molecule is not favored.
This conclusion is consistent with the prediction from the QCD sum
rule \cite{Lee:2008gn}. We suggest that the $1^{--}$, $2^{++}$,
$0^{++}$ and $0^{-+}$ $D^{*}\bar{D}^{*}$ molecules should very
likely exist, whereas the CP exotic $1^{-+}$ and $2^{+-}$
$D^{*}\bar{D}^{*}$ states, $2^{-+}$ and $2^{--}$ $D^{*}\bar{D}^{*}$
states should not be bound by the one boson exchange potential.
Although $Y(4008)$ is close to the $D^{*}\bar{D}^{*}$ threshold, its
width is so large that it is not reasonable to identify $Y(4008)$ as
a $D^{*}\bar{D}^{*}$ molecule. The $1^{--}$ $D^{*}\bar{D}^{*}$
molecular state can be produced copiously in $e^{+}e^{-}$
annihilation or via the initial state radiation at $B$ factory,
detailed scan of the $e^{+}e^{-}$ annihilation data near the
$D^{*}\bar{D}^{*}$ threshold is crucial to confirming this
prediction.

\section{The molecular states of the $B^{*}\bar{B}^{*}$ system}

The mass difference between $B^{*+}$ and $B^{*0}$ is so small that
it can be negligible \cite{pdg}. If we take into account the mass
splitting within the exchanged pion isospin multiplet, we should
solve similar coupled channel problems as the $D^{*}\bar{D}^{*}$
case. The repulsive kinetic energy is greatly reduced due to the
larger mass of $B^{*}$ meson, therefore the $B^{*}\bar{B}^{*}$
system should be more deeply bound than the $D^{*}\bar{D}^{*}$
system. Numerically solving the corresponding Schr${\ddot{\rm
o}}$dinger equation, we notice that the molecular states bound for
reasonable $\Lambda$ value have definite isospin, and the $I=0$
configuration is obviously much easier to be bound than the $I=1$
configuration. We also find that the binding energy dependence on
$\Lambda$ becomes less sensitive, if we reduce all the coupling
constants except $g_{\pi NN}$ by half. In the case that the mass
difference between $\pi^{\pm}$ and $\pi^{0}$ is neglected, the
$B^{*}\bar{B}^{*}$ state is of definite isospin, and the dimensions
of the coupled channel equations would be reduced by half. We have
seriously calculated the binding energy and the static properties
for both the $I=0$ and $I=1$ states. However, the numerical results
are too lengthy to be listed in the manuscript. We find that
introducing the pion mass splitting will modify the binding energy
by at most 0.5 MeV.

For the $0^{++}$ and $1^{+-}$ $B^{*}\bar{B}^{*}$ system, both the
$I=0$ and $I=1$ bound states can be found for the same value of the
regularization parameter $\Lambda$, whereas the two states behave in
different way. The $I=0$ state is generally more deeply bound than
the $I=1$ state. For the isospin singlet, the D wave components
increase drastically with $\Lambda$. Whereas for the isospin vector,
the S wave components are dominant, and they increase slightly with
$\Lambda$. Because $B^{*}$ mainly decays into $B\gamma$ \cite{pdg},
$BB\gamma\gamma$ is the leading decay mode of the $B^{*}\bar{B}^{*}$
molecule via the $B^{*}$ and $\bar{B}^{*}$ dissociation. For the
eight remaining states, we also find that the $I=0$ configuration is
more tightly bound than the $I=1$ configuration, hence it seems to
be a universal result that the hadronic molecule prefers to a
isospin singlet.

\begin{figure}[hptb]
\begin{center}
\begin{tabular}{ccc}
\includegraphics[scale=.645]{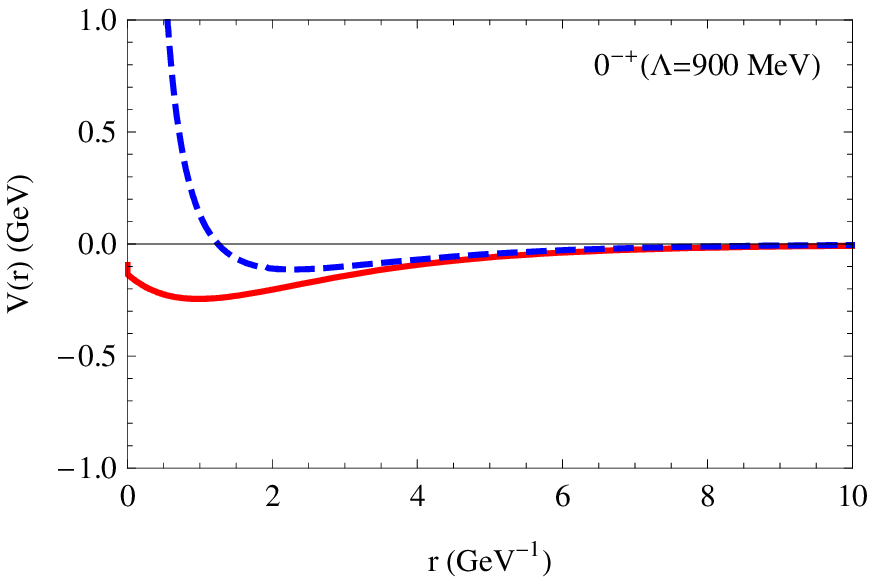}&\includegraphics[scale=.645]{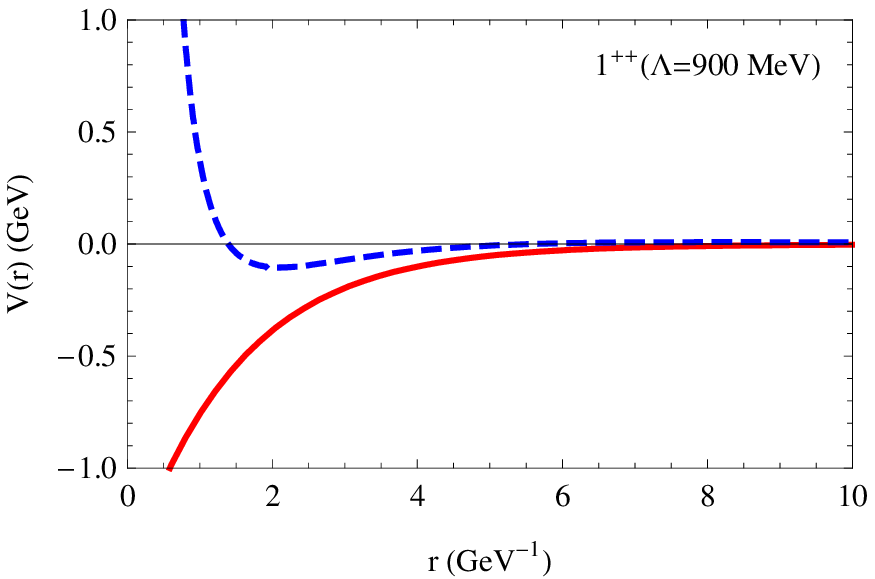}&\includegraphics[scale=.645]{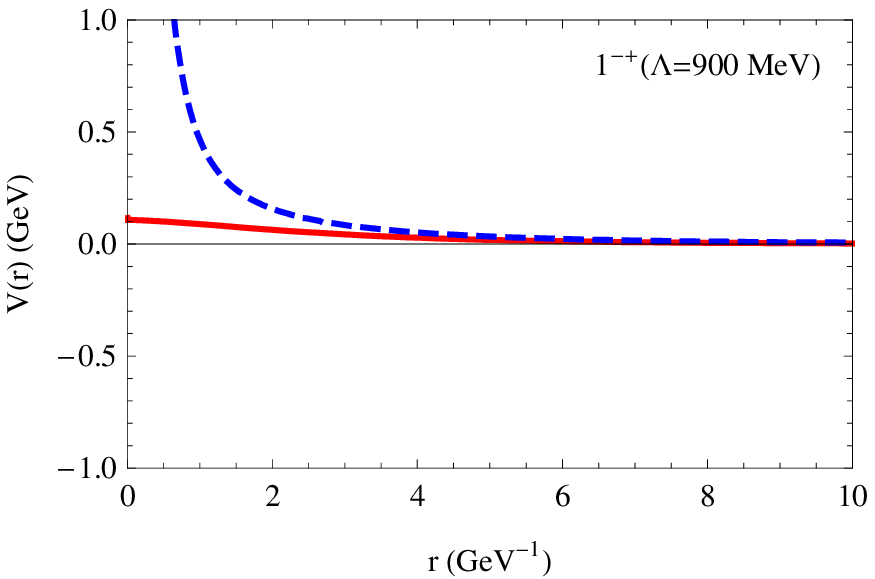}\\
(a)&(b)&(c)\\
\includegraphics[scale=.645]{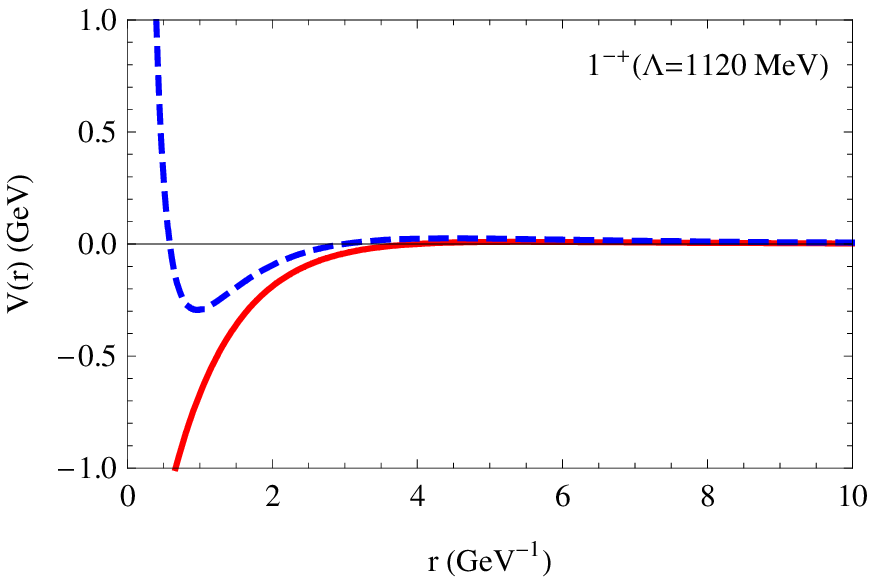}&\includegraphics[scale=.645]{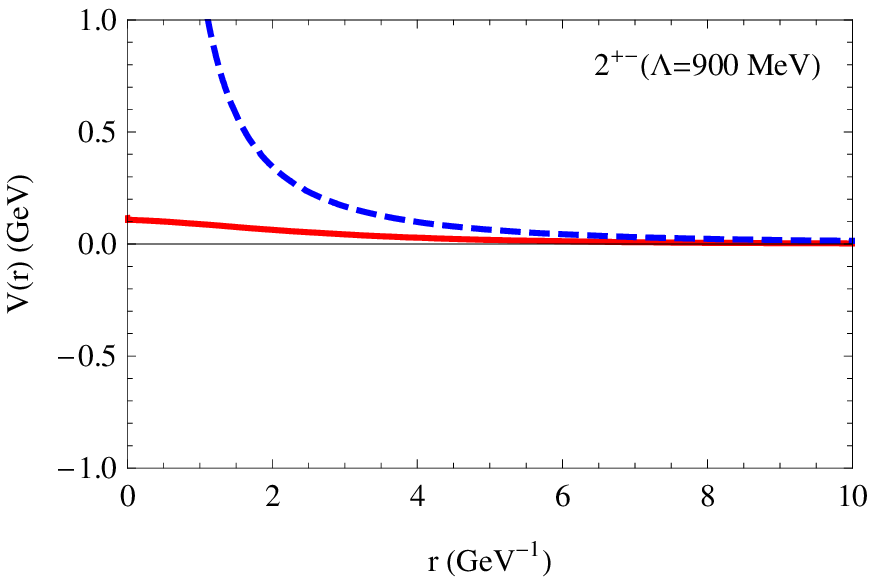}&\includegraphics[scale=.645]{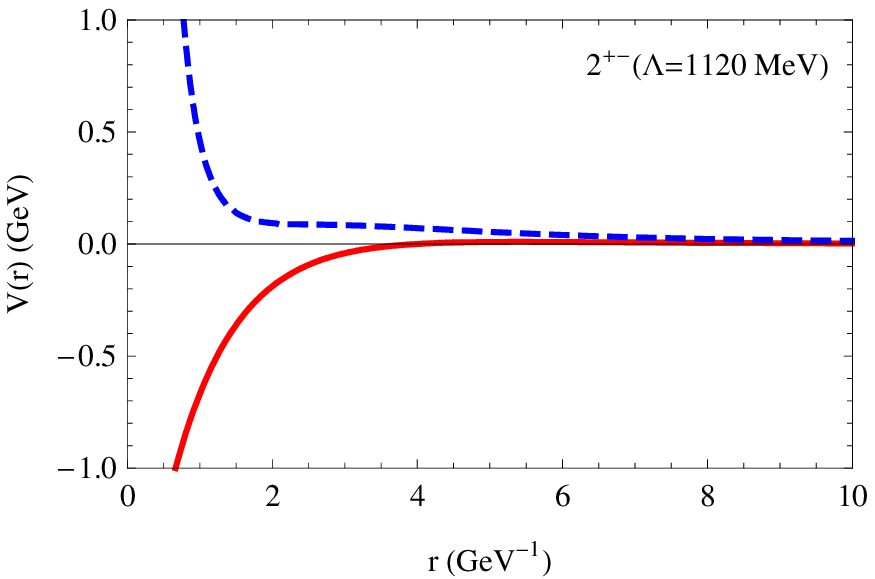}\\
(d)&(e)&(f)\\
\end{tabular}
\caption{\label{single_channel_potential}The potentials for the
single channel $0^{-+}$, $1^{++}$, $1^{-+}$ and $2^{+-}$
$B^{*}\bar{B}^{*}$ states with $I=0$. (a) and (b) show the
potentials for the $0^{-+}$ and $1^{++}$ states with $\Lambda=900$
MeV respectively. (c) and (d) are the potentials of the $1^{-+}$
states with $\Lambda=900$ MeV and 1120 MeV respectively. (e) and (f)
are for the $2^{+-}$ states with $\Lambda=900$ MeV and 1120 MeV
respectively. The solid line represents the potential from one boson
exchange, and the dashed line denotes the total potential including
the centrifugal barrier.}
\end{center}
\end{figure}

Since the mass splitting within the pion isospin multiplet only
introduces minor modifications. The numerical results can be
understood easily in the exact isospin symmetry limit. If the pion
mass splitting is neglected, the $0^{-+}$, $1^{++}$, $1^{-+}$ and
$2^{+-}$ $B^{*}\bar{B}^{*}$ states only involve one channel. Both
the one boson exchange potential and the total potential including
the centrifugal barrier for the four states are displayed in Fig.
\ref{single_channel_potential}. From Fig.
\ref{single_channel_potential}a, we can see that the P wave
centrifugal barrier is partly compensated by the one boson exchange
potential, and there remains a weak attractive interaction in the
intermediate range. Therefore $0^{-+}$ bound state can be found for
reasonable values of $\Lambda$. For $\Lambda=808-950$ MeV, we find
the binding energy is in the range of 2.93 to 134.71 MeV. Because
the D wave centrifugal barrier is higher than the P wave centrifugal
barrier, the total potential of the $1^{++}$ state is less
attractive than the $0^{-+}$ one for the same values of parameters,
this point can be seen clearly by comparing Fig.
\ref{single_channel_potential}b with Fig.
\ref{single_channel_potential}a. Accordingly our numerical results
really indicate that the $1^{++}$ $B^{*}\bar{B}^{*}$ state is harder
to be bound than the $0^{-+}$ state. For $\Lambda=900$ MeV, $1^{-+}$
$B^{*}\bar{B}^{*}$ bound state can not be found, it is because the
potential including the P wave centrifugal barrier is repulsive in
this case. When $\Lambda$ is increased to about 1120 MeV, the total
potential shown in Fig. \ref{single_channel_potential}d becomes
attractive in the intermediate region, the $1^{-+}$
$B^{*}\bar{B}^{*}$ system is marginally bound. The one boson
exchange potentials for the CP exotic $1^{-+}$ and $2^{+-}$ states
are exactly the same. However, the $2^{+-}$ $B^{*}\bar{B}^{*}$
system is more loosely bound than the $1^{-+}$ system because of the
D wave centrifugal barrier. Fig. \ref{single_channel_potential}e and
Fig. \ref{single_channel_potential}f clearly show that the total
potential of the $2^{+-}$ state is still repulsive for both
$\Lambda=900$ MeV and 1120 MeV. $2^{+-}$ $B^{*}\bar{B}^{*}$ bound
state begins to appear for $\Lambda$ as large as 1260 MeV.

\begin{figure}[hptb]
\begin{center}
\begin{tabular}{cccc}
\includegraphics[scale=.45]{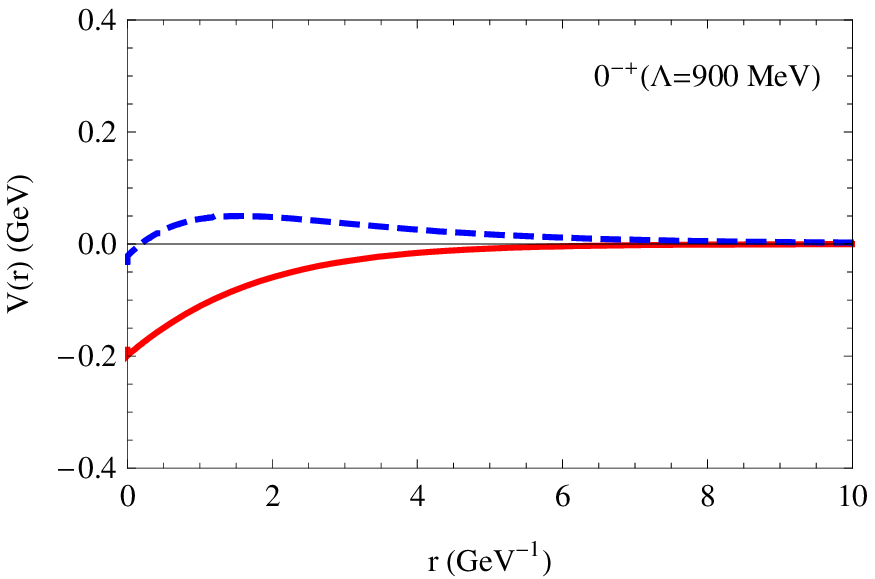}&\includegraphics[scale=.45]{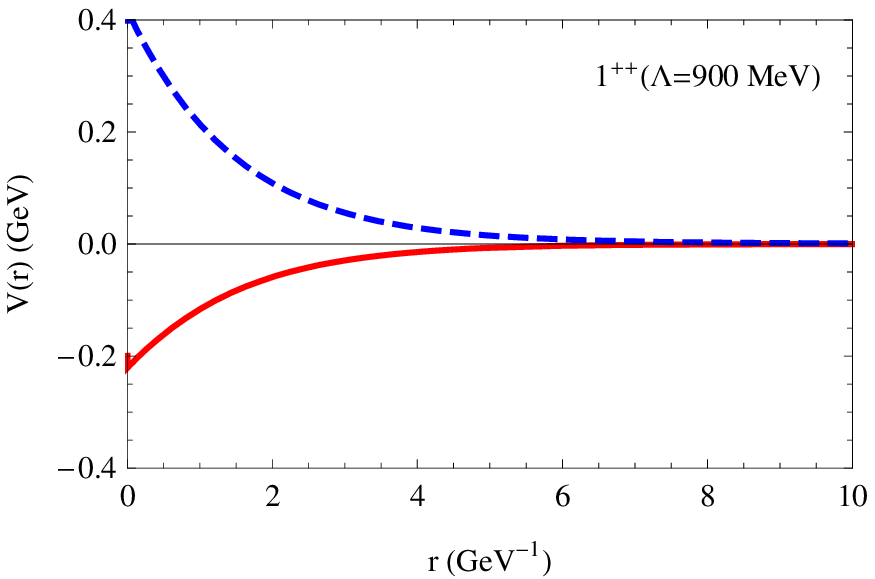}&\includegraphics[scale=.45]{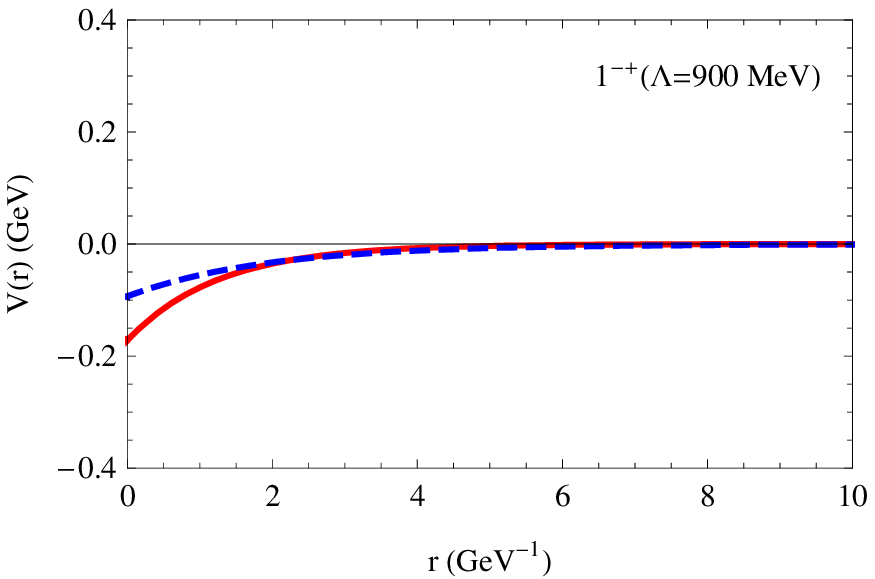}&\includegraphics[scale=.45]{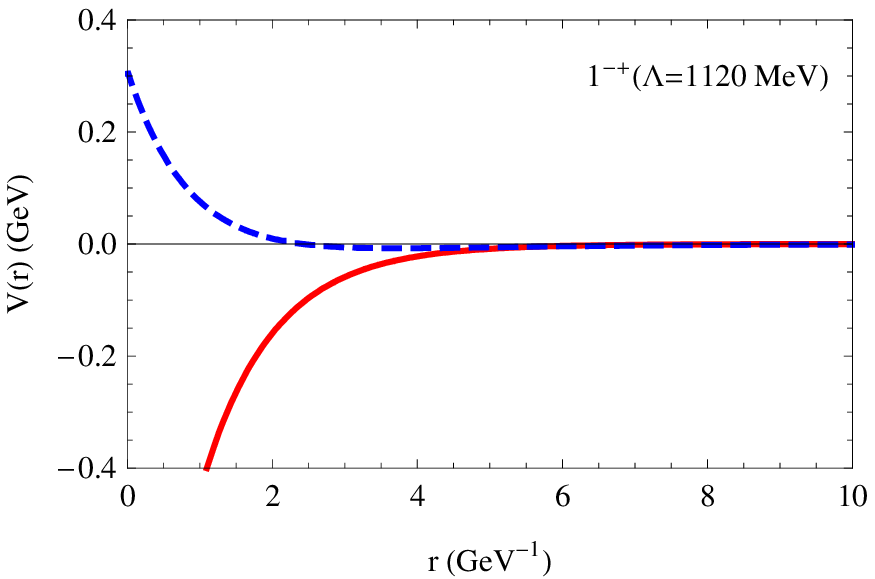}\\
(a)&(b)&(c)&(d)
\end{tabular}
\caption{\label{spin_potential}The isospin relevant and irrelevant
one boson exchange potentials for the $0^{-+}$, $1^{++}$ and
$1^{-+}$ $B^{*}\bar{B}^{*}$ states. (a) and (b) respectively show
the potentials for the $0^{-+}$ and $1^{++}$ states with
$\Lambda=900$ MeV. (c) and (d) are the potentials for the $1^{-+}$
states with $\Lambda=900$ MeV and 1120 MeV respectively. The one
boson exchange potential for the $2^{+-}$ state is exactly the same
as the $1^{-+}$ case. The solid line represents the isospin
irrelevant potential, and the dashed line denotes the isospin
relevant part.}
\end{center}
\end{figure}

The dominance of the $I=0$ configuration over $I=1$ can be clearly
understood as well for the $0^{-+}$, $1^{++}$, $1^{-+}$ and $2^{+-}$
states in the isospin symmetry limit. In this case, the one boson
exchange potential is the summation of the isospin independent part
and $C_I$ multiplying the isospin relevant part, where the parameter
$C_I$ is equal to -3 and 1 respectively for $I=0$ and 1. Concretely
for the $0^{-+}$ state, the isospin independent potential is
$V_C(r)-V_{S}(r)-2V_T(r)-V_{LS}(r)$, and the isospin relevant part
is
$V_I(\mu_{\rho},r)-V_{SI}(m_{\pi},\mu_{\pi},\mu_{\rho},r)-2V_{TI}(m_{\pi},\mu_{\pi},\mu_{\rho},r)-V_{LSI}(\mu_{\rho},r)$.
We plot both the isospin irrelevant and relevant potentials for
these states in Fig. \ref{spin_potential}, it is obvious that the
isospin irrelevant potential is usually attractive. For the $0^{-+}$
and $1^{++}$ states with $\Lambda=900$ MeV, the isospin relevant
potential is positive, hence it is easily understood why the $I=0$
configuration is more attractive than $I=1$. For the $1^{-+}$ state
with $\Lambda=900$ MeV, the isospin relevant part is negative, thus
the potential for the $I=1$ state is deeper than the one for $I=0$.
However, the corresponding potential still can not support a
$1^{-+}$ $B^{*}\bar{B}^{*}$ isovector state, such state can be bound
only when $\Lambda$ is increased to about $1200$ MeV. For $\Lambda$
as large as 1120 MeV, the isospin relevant potential becomes
positive, accordingly the $I=0$ configuration is more tightly bound
than $I=1$. In this case, a marginally bound $1^{-+}$
$B^{*}\bar{B}^{*}$ isospin singlet appears with mass about 10635.39
MeV. If we reduce all the coupling constants except $g_{\pi NN}$ by
half, the $1^{-+}$ and $2^{+-}$ $B^{*}\bar{B}^{*}$ states can be
bound only if $\Lambda$ is larger than 1700 MeV and $2790$ MeV
respectively. Therefore the one boson exchange potential may not
support the CP exotic $1^{-+}$ and $2^{+-}$ $B^{*}\bar{B}^{*}$
molecules.

Following the same methhod as the $D^{*}\bar{D}^{*}$ case and
carefully examining the numerical results, we suggest that the
$1^{--}$, $2^{++}$, $0^{++}$ and $0^{-+}$ $B^{*}\bar{B}^{*}$
molecules should exist, and the $1^{+-}$ $B^{*}\bar{B}^{*}$ bound
state also very likely exists. Similar to the $1^{--}$
$D^{*}\bar{D}^{*}$ state, the $1^{--}$ $B^{*}\bar{B}^{*}$ molecule
can be produced largely in $e^{+}e^{-}$ annihilation at Babar or
Belle. Detailed scanning of the $e^{+}e^{-}$ annihilation data at
the $B^{*}\bar{B}^{*}$ threshold is expected.

\section{Conclusions and discussions}

Motivated by the charmonium-like state $Y(4008)$ and
$Z^{+}_1(4050)$, the possible $D^{*}\bar{D}^{*}$ molecular states
have been studied dynamically in the one boson exchange model, where
$\pi$, $\eta$, $\sigma$, $\rho$ and $\omega$ exchanges are taken
into account. Ten allowed states with low spin parity have been
considered. We find that the binding energy and static properties
are sensitive to the regularization parameter $\Lambda$ and the
effective coupling constants. The binding energy increases with
$\Lambda$, whereas the root of mean square radius decreases with
$\Lambda$, this is because increasing $\Lambda$ increases the
strength of the potential at short distance. Larger coupling
constants are favorable to the formation of molecular states. If all
the coupling constants except $g_{\pi NN}$ are reduced by half,
larger value of $\Lambda$ is required to find bound state solutions.
Isospin violation is expected, especially for the states close to
the threshold, and the $I=0$ component is dominant, The predominance
of the $I=0$ configuration over $I=1$ can be clearly understood in
the exact isospin symmetry limit, and the same conclusion is reached
in the one pion exchange model. Hence the interpretation of
$Z^{+}_1(4050)$ as a $D^{*}\bar{D}^{*}$ molecule is not favored.

Since the regularization parameter $\Lambda$ is poorly known so far,
we are not be able to precisely predict the binding energies for the
possible molecular states bound by one boson exchange potential.
Certainly, if the potential is strong enough one can be quite
confident that such bound state must exist, but their exact binding
energy always depends on the details of the regularization
procedure. However, we can reliably predict which ones of the ten
allowed states are much easier to be bound, and the prediction is
rather stable even if the uncertainty of the coupling constants is
considered, as is obvious from the numerical results in Appendix
\ref{charm_result}. Further research on X(3872) would put severe
constraint on the parameters on the one boson exchange model,
especially on the regularization parameter $\Lambda$, so that the
predictions presented in the work could become more precise.

Our detailed numerical results indicate that the $1^{--}$, $2^{++}$,
$0^{++}$ and $0^{-+}$ $D^{*}\bar{D}^{*}$ bound states should very
likely exist, whereas the CP exotic $1^{-+}$ and $2^{+-}$
$D^{*}\bar{D}^{*}$ states, $2^{-+}$ and $2^{--}$ $D^{*}\bar{D}^{*}$
states may not be bound by the one boson exchange potential. The
$1^{--}$ state can be directly produced in $e^{+}e^{-}$
annihilation, detailed $e^{+}e^{-}$ annihilation data near the
$D^{*}\bar{D}^{*}$ threshold are important to confirm or refute the
existence of such state. The $D^{*}\bar{D}^{*}$ molecule mainly
decays into $D\bar{D}\gamma\gamma$ and $D\bar{D}\gamma\pi$ via the
dissociation of $D^{*}$ and $\bar{D}^{*}$, the $D\bar{D}\pi\pi$ mode
is highly suppressed or forbidden by the phase space, and its width
should be about tens of MeV. Although $Y(4008)$ is close to the
$D^{*}\bar{D}^{*}$ threshold, its width is huge so that it is
unreasonable to identify $Y(4008)$ as a $D^{*}\bar{D}^{*}$ molecule.
The existence of $Y(4008)$ and $Z^{+}_1(4050)$ has not been
confirmed so far, more experimental efforts are urgently needed.

The possible $B^{*}\bar{B}^{*}$ molecular states have been discussed
along the same line. The $B^{*}\bar{B}^{*}$ system is more deeply
bound than the $D^{*}\bar{D}^{*}$ system. We find that the molecular
states of the $B^{*}\bar{B}^{*}$ system have definite isospin, and
the isospin singlet is much easier to be bound than the isospin
vector for the states considered, the reason is analyzed in the
isospin symmetry limit. If the small isospin mass splitting between
the neutral and charged pion mesons is neglected, the dimension of
the coupled channel problem is reduced by half, the situation is
simplified greatly. It is observed that including the pion mass
splitting modifies the binding energy by at most 0.5 MeV. We suggest
that the $1^{--}$, $2^{++}$, $0^{++}$ and $0^{-+}$
$B^{*}\bar{B}^{*}$ molecules should exist, they would be narrow
states, and they dominantly decay into $B\bar{B}\gamma\gamma$.

\begin{acknowledgments}
We are grateful to Prof. Mu-Lin Yan and Dao-Neng Gao for stimulating
discussions. This work is supported by the China Postdoctoral
Science Foundation (20070420735), K.C. Wong Education Foundation,
and KJCX2-YW-N29 of the Chinese Academy.
\end{acknowledgments}

\newpage

\begin{appendix}

\section{The one boson exchange potential for the $D^{*}\bar{D}^{*}$($B^{*}\bar{B}^{*}$) states including the isospin mass splitting}

\begin{eqnarray}
\nonumber
&&V_{1^{+-}}=\big[V_{C}(r)-V_S(r)\big]\left(\begin{array}{cccc}
1&0&0&0\\
0&1&0&0\\
0&0&1&0\\
0&0&0&1
\end{array}\right)+\big[-V_{I}(\mu_{\rho},r)+V_{SI}(m_{\pi},\mu_{\pi},\mu_{\rho},r)\big]\left(\begin{array}{cccc}
1&0&2&0\\
0&1&0&2\\
2&0&1&0\\
0&2&0&1
\end{array}\right)\\
\nonumber&&+V_{T}(r)\left(\begin{array}{cccc}
0&\sqrt{2}&0&0\\
\sqrt{2}&-1&0&0\\
0&0&0&\sqrt{2}\\
0&0&\sqrt{2}&-1
\end{array}\right)+V_{TI}(m_{\pi},\mu_{\pi},\mu_{\rho},r)\left(\begin{array}{cccc}
0&-\sqrt{2}&0&-2\sqrt{2}\\
-\sqrt{2}&1&-2\sqrt{2}&2\\
0&-2\sqrt{2}&0&-\sqrt{2}\\
-2\sqrt{2}&2&-\sqrt{2}&1
\end{array}\right)\\
\label{a1}&&+V_{LS}(r)\left(\begin{array}{cccc}
0&0&0&0\\
0&-\frac{3}{2}&0&0\\
0&0&0&0\\
0&0&0&-\frac{3}{2}
\end{array}\right)+V_{LSI}(\mu_{\rho},r)\left(\begin{array}{cccc}
0&0&0&0\\
0&\frac{3}{2}&0&3\\
0&0&0&0\\
0&3&0&\frac{3}{2}
\end{array}\right)
\end{eqnarray}

\begin{eqnarray}
\nonumber
&&V_{0^{-+}}(r)=\big[V_{C}(r)-V_{S}(r)-2V_T(r)-V_{LS}(r)\big]\left(\begin{array}{cc}
1&0\\
0&1
\end{array}\right)+\big[-V_{I}(\mu_{\rho},r)+V_{SI}(m_{\pi},\mu_{\pi},\mu_{\rho},r)\\
\label{a2}&&+2V_{TI}(m_{\pi},\mu_{\pi},\mu_{\rho},r)+V_{LSI}(\mu_{\rho},r)\big]\left(\begin{array}{cc}
1&2\\
2&1
\end{array}
\right)
\end{eqnarray}

\begin{eqnarray}
\nonumber&&V_{1^{++}}(r)=\big[V_{C}(r)+V_{S}(r)-V_T(r)-\frac{5}{2}V_{LS}(r)\big]\left(\begin{array}{cc}
1&0\\
0&1
\end{array}\right)+\big[-V_{I}(\mu_{\rho},r)-V_{SI}(m_{\pi},\mu_{\pi},\mu_{\rho},r)\\
\label{a3}&&+V_{TI}(m_{\pi},\mu_{\pi},\mu_{\rho},r)+\frac{5}{2}V_{LSI}(\mu_{\rho},r)\big]\left(\begin{array}{cc}
1&2\\
2&1
\end{array}
\right)
\end{eqnarray}

\begin{eqnarray}
\nonumber&&V_{1^{-+}}(r)=\big[V_{C}(r)-V_{S}(r)+V_T(r)-\frac{1}{2}V_{LS}(r)\big]\left(\begin{array}{cc}
1&0\\
0&1
\end{array}\right)+\big[-V_{I}(\mu_{\rho},r)+V_{SI}(m_{\pi},\mu_{\pi},\mu_{\rho},r)\\
\label{a4}&&-V_{TI}(m_{\pi},\mu_{\pi},\mu_{\rho},r)+\frac{1}{2}V_{LSI}(\mu_{\rho},r)\big]\left(\begin{array}{cc}
1&2\\
2&1
\end{array}
\right)
\end{eqnarray}

\begin{eqnarray}
\nonumber&&V_{2^{+-}}(r)=\big[V_{C}(r)-V_{S}(r)+V_T(r)-\frac{1}{2}V_{LS}(r)\big]\left(\begin{array}{cc}
1&0\\
0&1
\end{array}\right)+\big[-V_{I}(\mu_{\rho},r)+V_{SI}(m_{\pi},\mu_{\pi},\mu_{\rho},r)\\
\label{a5}&&-V_{TI}(m_{\pi},\mu_{\pi},\mu_{\rho},r)+\frac{1}{2}V_{LSI}(\mu_{\rho},r)\big]\left(\begin{array}{cc}
1&2\\
2&1
\end{array}
\right)
\end{eqnarray}

\begin{eqnarray}
\nonumber&&V_{1^{--}}(r)=V_{C}(r)\left(\begin{array}{cccccc}
1&0&0&0&0&0\\
0&1&0&0&0&0\\
0&0&1&0&0&0\\
0&0&0&1&0&0\\
0&0&0&0&1&0\\
0&0&0&0&0&1\\
\end{array}\right)+V_{S}(r)\left(\begin{array}{cccccc}
-2&0&0&0&0&0\\
0&1&0&0&0&0\\
0&0&1&0&0&0\\
0&0&0&-2&0&0\\
0&0&0&0&1&0\\
0&0&0&0&0&1\\
\end{array}\right)+V_{I}(\mu_{\rho},r)\left(\begin{array}{cccccc}
-1&0&0&-2&0&0\\
0&-1&0&0&-2&0\\
0&0&-1&0&0&-2\\
-2&0&0&-1&0&0\\
0&-2&0&0&-1&0\\
0&0&-2&0&0&-1\\
\end{array}\right)\\
\nonumber&&+V_{T}(r)\left(\begin{array}{cccccc}
0&\frac{2}{\sqrt{5}}&-\sqrt{\frac{6}{5}}&0&0&0\\
\frac{2}{\sqrt{5}}&-\frac{7}{5}&\frac{\sqrt{6}}{5}&0&0&0\\
-\sqrt{\frac{6}{5}}&\frac{\sqrt{6}}{5}&-\frac{8}{5}&0&0&0\\
0&0&0&0&\frac{2}{\sqrt{5}}&-\sqrt{\frac{6}{5}}\\
0&0&0&\frac{2}{\sqrt{5}}&-\frac{7}{5}&\frac{\sqrt{6}}{5}\\
0&0&0&-\sqrt{\frac{6}{5}}&\frac{\sqrt{6}}{5}&-\frac{8}{5}\\
\end{array}\right)+V_{SI}(m_{\pi},\mu_{\pi},\mu_{\rho},r)\left(\begin{array}{cccccc}
2&0&0&4&0&0\\
0&-1&0&0&-2&0\\
0&0&-1&0&0&-2\\
4&0&0&2&0&0\\
0&-2&0&0&-1&0\\
0&0&-2&0&0&-1\\
\end{array}\right)\\
\nonumber&&+V_{TI}(m_{\pi},\mu_{\pi},\mu_{\rho},r)\left(\begin{array}{cccccc}
0&-\frac{2}{\sqrt{5}}&\sqrt{\frac{6}{5}}&0&-\frac{4}{\sqrt{5}}&2\sqrt{\frac{6}{5}}\\
-\frac{2}{\sqrt{5}}&\frac{7}{5}&-\frac{\sqrt{6}}{5}&-\frac{4}{\sqrt{5}}&\frac{14}{5}&-\frac{2\sqrt{6}}{5}\\
\sqrt{\frac{6}{5}}&-\frac{\sqrt{6}}{5}&\frac{8}{5}&2\sqrt{\frac{6}{5}}&-\frac{2\sqrt{6}}{5}&\frac{16}{5}\\
0&-\frac{4}{\sqrt{5}}&2\sqrt{\frac{6}{5}}&0&-\frac{2}{\sqrt{5}}&\sqrt{\frac{6}{5}}\\
-\frac{4}{\sqrt{5}}&\frac{14}{5}&-\frac{2\sqrt{6}}{5}&-\frac{2}{\sqrt{5}}&\frac{7}{5}&-\frac{\sqrt{6}}{5}\\
2\sqrt{\frac{6}{5}}&-\frac{2\sqrt{6}}{5}&\frac{16}{5}&\sqrt{\frac{6}{5}}&-\frac{\sqrt{6}}{5}&\frac{8}{5}
\end{array}\right)+V_{LS}(r)\left(\begin{array}{cccccc}
0&0&0&0&0&0\\
0&-\frac{3}{2}&0&0&0&0\\
0&0&-4&0&0&0\\
0&0&0&0&0&0\\
0&0&0&0&-\frac{3}{2}&0\\
0&0&0&0&0&-4\\
\end{array}\right)\\
\label{a6}&&+V_{LSI}(\mu_{\rho},r)\left(\begin{array}{cccccc}
0&0&0&0&0&0\\
0&\frac{3}{2}&0&0&3&0\\
0&0&4&0&0&8\\
0&0&0&0&0&0\\
0&3&0&0&\frac{3}{2}&0\\
0&0&8&0&0&4\\
\end{array}\right)
\end{eqnarray}

\begin{eqnarray}
\nonumber&&V_{2^{++}}=V_{C}(r)\left(\begin{array}{cccccccc}
1&0&0&0&0&0&0&0\\
0&1&0&0&0&0&0&0\\
0&0&1&0&0&0&0&0\\
0&0&0&1&0&0&0&0\\
0&0&0&0&1&0&0&0\\
0&0&0&0&0&1&0&0\\
0&0&0&0&0&0&1&0\\
0&0&0&0&0&0&0&1
\end{array}
\right)+V_{S}(r)\left(\begin{array}{cccccccc}
-2&0&0&0&0&0&0&0\\
0&1&0&0&0&0&0&0\\
0&0&1&0&0&0&0&0\\
0&0&0&1&0&0&0&0\\
0&0&0&0&-2&0&0&0\\
0&0&0&0&0&1&0&0\\
0&0&0&0&0&0&1&0\\
0&0&0&0&0&0&0&1
\end{array}
\right)\\
\nonumber&&+V_{I}(\mu_{\rho},r)\left(\begin{array}{cccccccc}
-1&0&0&0&-2&0&0&0\\
0&-1&0&0&0&-2&0&0\\
0&0&-1&0&0&0&-2&0\\
0&0&0&-1&0&0&0&-2\\
-2&0&0&0&-1&0&0&0\\
0&-2&0&0&0&-1&0&0\\
0&0&-2&0&0&0&-1&0\\
0&0&0&-2&0&0&0&-1
\end{array}
\right)+V_{T}(r)\left(\begin{array}{cccccccc}
0&-\sqrt{\frac{2}{5}}&\frac{2}{\sqrt{7}}&-\frac{6}{\sqrt{35}}&0&0&0&0\\
-\sqrt{\frac{2}{5}}&0&\sqrt{\frac{14}{5}}&0&0&0&0&0\\
\frac{2}{\sqrt{7}}&\sqrt{\frac{14}{5}}&\frac{3}{7}&\frac{12}{7\sqrt{5}}&0&0&0&0\\
-\frac{6}{\sqrt{35}}&0&\frac{12}{7\sqrt{5}}&-\frac{10}{7}&0&0&0&0\\
0&0&0&0&0&-\sqrt{\frac{2}{5}}&\frac{2}{\sqrt{7}}&-\frac{6}{\sqrt{35}}\\
0&0&0&0&-\sqrt{\frac{2}{5}}&0&\sqrt{\frac{14}{5}}&0\\
0&0&0&0&\frac{2}{\sqrt{7}}&\sqrt{\frac{14}{5}}&\frac{3}{7}&\frac{12}{7\sqrt{5}}\\
0&0&0&0&-\frac{6}{\sqrt{35}}&0&\frac{12}{7\sqrt{5}}&-\frac{10}{7}
\end{array}
\right)\\
\nonumber&&+V_{SI}(m_{\pi},\mu_{\pi},\mu_{\rho},r)\left(\begin{array}{cccccccc}
2&0&0&0&4&0&0&0\\
0&-1&0&0&0&-2&0&0\\
0&0&-1&0&0&0&-2&0\\
0&0&0&-1&0&0&0&-2\\
4&0&0&0&2&0&0&0\\
0&-2&0&0&0&-1&0&0\\
0&0&-2&0&0&0&-1&0\\
0&0&0&-2&0&0&0&-1
\end{array}
\right)\\
\nonumber&&+V_{TI}(m_{\pi},\mu_{\pi},\mu_{\rho},r)\left(\begin{array}{cccccccc}
0&\sqrt{\frac{2}{5}}&-\frac{2}{\sqrt{7}}&\frac{6}{\sqrt{35}}&0&2\sqrt{\frac{2}{5}}&-\frac{4}{\sqrt{7}}&\frac{12}{\sqrt{35}}\\
\sqrt{\frac{2}{5}}&0&-\sqrt{\frac{14}{5}}&0&2\sqrt{\frac{2}{5}}&0&-2\sqrt{\frac{14}{5}}&0\\
-\frac{2}{\sqrt{7}}&-\sqrt{\frac{14}{5}}&-\frac{3}{7}&-\frac{12}{7\sqrt{5}}&-\frac{4}{\sqrt{7}}&-2\sqrt{\frac{14}{5}}&-\frac{6}{7}&-\frac{24}{7\sqrt{5}}\\
\frac{6}{\sqrt{35}}&0&-\frac{12}{7\sqrt{5}}&\frac{10}{7}&\frac{12}{\sqrt{35}}&0&-\frac{24}{7\sqrt{5}}&\frac{20}{7}\\
0&2\sqrt{\frac{2}{5}}&-\frac{4}{\sqrt{7}}&\frac{12}{\sqrt{35}}&0&\sqrt{\frac{2}{5}}&-\frac{2}{\sqrt{7}}&\frac{6}{\sqrt{35}}\\
2\sqrt{\frac{2}{5}}&0&-2\sqrt{\frac{14}{5}}&0&\sqrt{\frac{2}{5}}&0&-\sqrt{\frac{14}{5}}&0\\
-\frac{4}{\sqrt{7}}&-2\sqrt{\frac{14}{5}}&-\frac{6}{7}&-\frac{24}{7\sqrt{5}}&-\frac{2}{\sqrt{7}}&-\sqrt{\frac{14}{5}}&-\frac{3}{7}&-\frac{12}{7\sqrt{5}}\\
\frac{12}{\sqrt{35}}&0&-\frac{24}{7\sqrt{5}}&\frac{20}{7}&\frac{6}{\sqrt{35}}&0&-\frac{12}{7\sqrt{5}}&\frac{10}{7}
\end{array}
\right)\\
\label{a7}&&+V_{LS}(r)\left(\begin{array}{cccccccc}
0&0&0&0&0&0&0&0\\
0&0&0&0&0&0&0&0\\
0&0&-\frac{3}{2}&0&0&0&0&0\\
0&0&0&-5&0&0&0&0\\
0&0&0&0&0&0&0&0\\
0&0&0&0&0&0&0&0\\
0&0&0&0&0&0&-\frac{3}{2}&0\\
0&0&0&0&0&0&0&-5
\end{array}
\right)+V_{LSI}(\mu_{\rho},r)\left(\begin{array}{cccccccc}
0&0&0&0&0&0&0&0\\
0&0&0&0&0&0&0&0\\
0&0&\frac{3}{2}&0&0&0&3&0\\
0&0&0&5&0&0&0&10\\
0&0&0&0&0&0&0&0\\
0&0&0&0&0&0&0&0\\
0&0&3&0&0&0&\frac{3}{2}&0\\
0&0&0&10&0&0&0&5
\end{array}
\right)
\end{eqnarray}

\begin{eqnarray}
\nonumber
&&V_{2^{-+}}(r)=\big[V_C(r)-V_S(r)\big]\left(\begin{array}{cccc}
1&0&0&0\\
0&1&0&0\\
0&0&1&0\\
0&0&0&1
\end{array}\right)+\big[-V_I(\mu_{\rho},r)+V_{SI}(m_{\pi},\mu_{\pi},\mu_{\rho},r)\big]\left(\begin{array}{cccc}
1&0&2&0\\
0&1&0&2\\
2&0&1&0\\
0&2&0&1
\end{array}\right)\\
\nonumber&&+V_T(r)\left(\begin{array}{cccc}
-\frac{1}{5}&\frac{3\sqrt{6}}{5}&0&0\\
\frac{3\sqrt{6}}{5}&-\frac{4}{5}&0&0\\
0&0&-\frac{1}{5}&\frac{3\sqrt{6}}{5}\\
0&0&\frac{3\sqrt{6}}{5}&-\frac{4}{5}
\end{array}\right)+V_{TI}(m_{\pi},\mu_{\pi},\mu_{\rho},r)\left(\begin{array}{cccc}
\frac{1}{5}&-\frac{3\sqrt{6}}{5}&\frac{2}{5}&-\frac{6\sqrt{6}}{5}\\
-\frac{3\sqrt{6}}{5}&\frac{4}{5}&-\frac{6\sqrt{6}}{5}&\frac{8}{5}\\
\frac{2}{5}&-\frac{6\sqrt{6}}{5}&\frac{1}{5}&-\frac{3\sqrt{6}}{5}\\
-\frac{6\sqrt{6}}{5}&\frac{8}{5}&-\frac{3\sqrt{6}}{5}&\frac{4}{5}\end{array}\right)\\
\label{a8}&&+V_{LS}(r)\left(\begin{array}{cccc}
\frac{1}{2}&0&0&0\\
0&-2&0&0\\
0&0&\frac{1}{2}&0\\
0&0&0&-2
\end{array}\right)+V_{LSI}(\mu_{\rho},r)\left(\begin{array}{cccc}
-\frac{1}{2}&0&-1&0\\
0&2&0&4\\
-1&0&-\frac{1}{2}&0\\
0&4&0&2
\end{array}\right)
\end{eqnarray}

\begin{eqnarray}
\nonumber&&V_{2^{--}}(r)=\big[V_C(r)+V_S(r)\big]\left(\begin{array}{cccc}
1&0&0&0\\
0&1&0&0\\
0&0&1&0\\
0&0&0&1
\end{array}
\right)+\big[V_I(\mu_{\rho},r)+V_{SI}(m_{\pi},\mu_{\pi},\mu_{\rho},r)\big]\left(\begin{array}{cccc}
-1&0&-2&0\\
0&-1&0&-2\\
-2&0&-1&0\\
0&-2&0&-1
\end{array}
\right)\\
\nonumber&&+V_{T}(r)\left(\begin{array}{cccc}
\frac{7}{5}&\frac{6}{5}&0&0\\
\frac{6}{5}&-\frac{2}{5}&0&0\\
0&0&\frac{7}{5}&\frac{6}{5}\\
0&0&\frac{6}{5}&-\frac{2}{5}
\end{array}
\right)+V_{TI}(m_{\pi},\mu_{\pi},\mu_{\rho},r)\left(\begin{array}{cccc}
-\frac{7}{5}&-\frac{6}{5}&-\frac{14}{5}&-\frac{12}{5}\\
-\frac{6}{5}&\frac{2}{5}&-\frac{12}{5}&\frac{4}{5}\\
-\frac{14}{5}&-\frac{12}{5}&-\frac{7}{5}&-\frac{6}{5}\\
-\frac{12}{5}&\frac{4}{5}&-\frac{6}{5}&\frac{2}{5}
\end{array}
\right)+V_{LS}(r)\left(\begin{array}{cccc}
-\frac{1}{2}&0&0&0\\
0&-3&0&0\\
0&0&-\frac{1}{2}&0\\
0&0&0&-3
\end{array}
\right)\\
\label{a9}&&+V_{LSI}(\mu_{\rho},r)\left(\begin{array}{cccc}
\frac{1}{2}&0&1&0\\
0&3&0&6\\
1&0&\frac{1}{2}&0\\
0&6&0&3
\end{array}
\right)
\end{eqnarray}
In the above expressions, the parameters $m_{\pi}$, $\mu_{\pi}$ and
$\mu_{\rho}$ could take two different sets of values. For
$D^{*0}\bar{D}^{*0}\rightarrow D^{*0}\bar{D}^{*0}$
($B^{*0}\bar{B}^{*0}\rightarrow B^{*0}\bar{B}^{*0}$) and
$D^{*+}D^{*-}\rightarrow D^{*+}D^{*-}$ ($B^{*+}B^{*-}\rightarrow
B^{*+}B^{*-}$), we should choose $m_{\pi}=m_{\pi^{0}}$,
$\mu_{\pi}=m_{\pi^{0}}$ and $\mu_{\rho}=m_{\rho}$. Whereas for the
processes $D^{*0}\bar{D}^{*0}\rightarrow D^{*+}D^{*-}$
($B^{*0}\bar{B}^{*0}\rightarrow B^{*+}B^{*-}$) and
$D^{*+}D^{*-}\rightarrow D^{*0}\bar{D}^{*0}$
($B^{*+}B^{*-}\rightarrow B^{*0}\bar{B}^{*0}$), we should take
$m_{\pi}=m_{\pi^{\pm}}$,
$\mu_{\pi}=[m^2_{\pi^{\pm}}-(m_{D^{*+}}-m_{D^{*0}})^2]^{1/2}$
($\mu_{\pi}=m_{\pi^{\pm}}$) and
$\mu_{\rho}=[m^2_{\rho}-(m_{D^{*+}}-m_{D^{*0}})^2]^{1/2}$
($\mu_{\rho}=m_{\rho}$) respectively.

\section{Numerical results for the $D^{*}\bar{D}^{*}$ states\label{charm_result}}

\begin{center}
\begin{table}[hptb]
\begin{tabular}{|c|ccc|}\hline\hline

$\Lambda({\rm MeV})$&$~~~{\rm M}(\rm MeV)$&$~~~{\rm r}_{\rm
rms}({\rm fm})$&$~~~{\rm
P^{00}_S:P^{00}_D:P^{+-}_S:P^{+-}_D(\%)}$\\\hline
930& 4013.80     &8.24    & 95.69:1.80:0.50:2.01    \\
940&  4011.72   &  2.39    &   70.69:11.01:7.43:10.88   \\
950& 4004.40    &   1.36 &  39.99:22.32:16.38:21.31   \\
960&   3990.26  & 1.03     &   25.21:29.74:16.55:28.50 \\
970&  3968.64  &  0.86    &  17.70:34.73:14.01:33.57   \\
980&   3938.98   &  0.76   &  13.19:38.20:11.41:37.19    \\
990&  3900.83   & 0.68     &  10.22:40.68:9.27:39.83    \\
\hline\hline

\multicolumn{4}{|c|}{all couplings are reduced by half except
$g_{\pi NN}$}\\\hline\hline

$\Lambda({\rm MeV})$&$~~~{\rm M}(\rm MeV)$&$~~~{\rm r}_{\rm
rms}({\rm fm})$&$~~~{\rm P^{00}_S:P^{00}_D:P^{+-}_S:P^{+-}_D}(\%)$\\
\hline
1100&  4011.77   &  2.36  & 64.96:12.17:10.77:12.10   \\
1130  & 3998.69    &  1.15  &30.86:26.00:18.08:25.06 \\
1160  &   3971.28   & 0.85   &18.83:33.55:15.04:32.59  \\
1190  &  3927.61   &   0.70   & 13.03:38.09:11.59:37.29 \\
1220  & 3866.06  &   0.61  &  9.66:40.98:9.01:40.35
\\ \hline \hline
\end{tabular}
\caption{\label{charm0++}The predictions for the static properties
of the ${\rm J^{PC}=0^{++}} $ ${\rm D^{*}\bar{D}^{*}}$ hadronic
molecule, where M denotes the mass, rms is the root of mean square
radius, ${\rm P_S}$ and ${\rm P_D}$ represent the S state and D
state probabilities respectively.}
\end{table}
\end{center}

\begin{center}
\begin{table}%[hptb]
\begin{tabular}{|c|ccc|}\hline\hline

$\Lambda({\rm MeV})$&$~~~{\rm M}(\rm MeV)$&$~~~{\rm r}_{\rm
rms}({\rm fm})$&~~~ ${\rm P^{00}_P:P^{+-}_P}(\%)$\\\hline

950 & 4011.33   &1.76   & 58.98:41.02 \\

970 &  3995.72  & 1.13 & 53.55:46.45 \\

990 &  3969.26  & 0.89 & 51.99:48.01 \\

1010 & 3930.16    &0.74   & 51.28:48.72 \\

1030 & 3876.73   & 0.64 &50.88:49.12  \\
\hline\hline

\multicolumn{4}{|c|}{all couplings except $g_{\pi NN}$ are reduced
by half}\\\hline\hline

$\Lambda({\rm MeV})$&$~~~{\rm M}(\rm MeV)$&$~~~{\rm r}_{\rm
rms}({\rm fm})$& ~~~${\rm P^{00}_P:P^{+-}_P}(\%)$\\
\hline

1120   &  4012.74     &  2.05   & 60.78:39.22  \\

1150  &  4004.34   &   1.30  &  54.77:45.23 \\

1200  & 3980.14    & 0.94    & 52.20:47.80 \\

1250  & 3940.92    &  0.75   &  51.28:48.72 \\

1300  &  3883.95   &  0.63   & 50.83:49.17  \\
\hline \hline

\end{tabular}
\caption{\label{charm0-+}The predictions about the mass, the root of
mean square radius(rms) and the probabilities of the different
components for the $0^{-+}$ ${\rm D^{*}\bar{D}^{*}}$ molecule.}
\end{table}
\end{center}

\begin{center}
\begin{table}%[hptb]
\begin{tabular}{|c|ccc|}\hline\hline

$\Lambda({\rm MeV})$&$~~~{\rm M}(\rm MeV)$&$~~~{\rm r}_{\rm
rms}({\rm fm})$&$~~~{\rm
P^{00}_{P0}:P^{00}_{P2}:P^{00}_{F}:P^{+-}_{P0}:P^{+-}_{P2}:P^{+-}_{F}(\%)}$\\\hline

920      & 4006.82     &  1.35    & 4.68:50.36:0.42:3.59:40.55:0.40   \\

930      &  3997.17   &  1.11     & 3.79:49.19:0.47:3.21:42.89:0.45    \\

940      &  3984.57  &    0.97  &   3.21:48.71:0.51:2.86:44.19:0.50   \\

950      & 3968.85    &  0.87   &   2.80:48.47:0.57:2.57:45.04:0.55    \\

960      &  3949.86    & 0.79    & 2.49:48.33:0.62:2.33:45.61:0.61     \\

970      &   3927.45 &  0.73   &  2.25:48.21:0.70:2.13:46.02:0.69   \\

980      &   3901.44 &  0.68   &   2.06:48.11:0.79:1.97:46.29:0.78  \\
\hline\hline

\multicolumn{4}{|c|}{all couplings except $g_{\pi NN}$ are reduced
by half}\\\hline\hline

$\Lambda({\rm MeV})$&$~~~{\rm M}(\rm MeV)$&$~~~{\rm r}_{\rm
rms}({\rm fm})$&$~~~{\rm P^{00}_{P0}:P^{00}_{P2}:P^{00}_{F}:P^{+-}_{P0}:P^{+-}_{P2}:P^{+-}_{F}}(\%)$\\
\hline

1050& 4007.15  &1.34  &5.36:49.41:0.49:4.16:40.10:0.48  \\

1080&  3991.87 &1.01 & 4.10:48.03:0.59:3.60:43.11:0.58 \\

1110&  3969.82  & 0.84 &3.38:47.66:0.67:3.12:44.51:0.66  \\

1140& 3940.40  & 0.73 & 2.90:47.52:0.77:2.75:45.31:0.75 \\

1170&  3903.00 &0.65  & 2.56:47.43:0.88:2.46:45.79:0.87 \\
\hline \hline

\end{tabular}
\caption{\label{charm1--}The predictions about the mass, the root of
mean square radius(rms) and the probabilities of the different
components for the $1^{--}$ ${\rm D^{*}\bar{D}^{*}}$ molecule, where
${\rm P^{00}_{P0}}$ and ${\rm P^{00}_{P2}}$ denote the ${\rm ^1P_1}$
and ${\rm ^5P_1}$ ${\rm D^{*0}\bar{D}^{*0}}$ states probabilities
respectively.}
\end{table}
\end{center}

\begin{center}
\begin{table}%[hptb]
\begin{tabular}{|c|ccc|}\hline\hline

$\Lambda({\rm MeV})$&$~~~{\rm M}(\rm MeV)$&$~~~{\rm r}_{\rm
rms}({\rm fm})$&$~~~{\rm
P^{00}_{D0}:P^{00}_{S}:P^{00}_{D2}:P^{00}_{G}:P^{+-}_{D0}:P^{+-}_{S}:P^{+-}_{D2}:P^{+-}_{G}(\%)}$\\\hline

860&   4010.82     & 1.96      &  0.39:67.73:1.95:0.01:0.38:27.59:1.96:0.01       \\

890&  4007.14     &  1.49    &    0.51:60.38:2.68:0.01:0.49:33.27:2.64:0.01    \\

920&  4001.88    &  1.26     &   0.64:55.35:3.60:0.02:0.61:36.24:3.51:0.02     \\

950&    3994.34  &   1.10    &    0.79:51.20:4.86:0.04:0.76:37.58:4.72:0.04   \\

980    &    3983.27 &   0.98    &   0.98:47.19:6.63:0.09:0.94:37.63:6.43:0.09    \\

1010    &   3966.58     & 0.88      &     1.21:42.85:9.11:0.22:1.17:36.37:8.85:0.22   \\

1040    &  3940.41     &   0.79   &    1.51:37.53:12.49:0.69:1.47:33.44:12.18:0.68   \\

1070    &  3894.04     & 0.69     &    2.01:29.16:16.55:3.55:1.96:27.02:16.22:3.53   \\
\hline\hline

\multicolumn{4}{|c|}{all couplings except $g_{\pi NN}$ are reduced
by half }\\\hline\hline

$\Lambda({\rm MeV})$&$~~~{\rm M}(\rm MeV)$&$~~~{\rm r}_{\rm
rms}({\rm fm})$&$~~~{\rm P^{00}_{D0}:P^{00}_{S}:P^{00}_{D2}:P^{00}_{G}:P^{+-}_{D0}:P^{+-}_{S}:P^{+-}_{D2}:P^{+-}_{G}}(\%)$\\
\hline

970& 4009.21    &  1.66   &   0.54:61.73:2.69:0.01:0.53:31.79:2.69:0.01   \\

1000&  4006.34   & 1.41    &   0.63:57.50:3.18:0.02:0.60:34.90:3.15:0.02   \\

1100&  3991.66  &  1.01   &    0.89:49.31:5.04:0.05:0.87:38.86:4.94:0.05 \\

1200& 3965.34   &  0.81   &   1.21:43.80:7.63:0.14:1.18:38.43:7.48:0.14   \\

1300& 3918.43     &  0.68   &  1.59:38.25:11.19:0.47:1.56:35.48:11.01:0.46  \\

1350&3882.03     &  0.62  &  1.82:34.98:13.32:0.95:1.79:33.04:13.15:0.95   \\
\hline \hline

\end{tabular}
\caption{\label{charm2++} The predictions about the mass, the root
of mean square radius(rms) and the probabilities of the different
components for the $2^{++}$ ${\rm D^{*}\bar{D}^{*}}$ molecule.}
\end{table}
\end{center}

\end{appendix}

\end{document}